\documentclass[10pt]{article}
\usepackage{multicol}
\usepackage{graphicx}
\usepackage{amsmath}
\usepackage[a4paper]{geometry}
\usepackage{hyperref}
\usepackage{rotating}

\begin{document}

\begin{center}

{\Large \bf Thermodynamic Signatures and Phase Transitions in High-Energy \vskip0.3cm
Au-Au Collisions} 

\vskip1.0cm

Murad Badshah$^{1}$,{\footnote{murad\_phy@awkum.edu.pk}}, 
Muhammad Waqas$^{2,}${\footnote{Corresponding author: 20220073@huat.edu.cn}},
Muhammad Ajaz$^{1,}${\footnote{Corresponding author: ajaz@awkum.edu.pk}}, 
Wolfgang Bietenholz$^{3}${\footnote{Corresponding author: wolbi@nucleares.unam.mx }}
Haifa I. Alrebdi$^{4,}$,{\footnote{Corresponding author: hialrebdi@pnu.edu.sa}}
Mohamed Ben Ammar$^{5,}$,%{\footnote{mohammed.ammar@nbu.edu.sa}}
\\
{\small\it $^1$ Department of Physics, Abdul Wali Khan University Mardan, Mardan 23200, Pakistan \\
$^2$ School of Mathematics, Physics and Optoelectronic Engineering, Hubei University of Automotive Technology 442002, Shiyan, People's Republic of China\\
$^3$ Instituto de Ciencias Nucleares - Universidad Nacional Autónoma de México, Apartado Postal 70-543, CdMx 04510, Mexico
\\
$^4$ Department of Physics, College of Science, Princess Nourah bint Abdulrahman University, P.O. Box 84428, Riyadh 11671, Saudi Arabia
\\
$^5$ Faculty of Computing and Information Technology, Northern Border University, Rafha, Saudi Arabia
}

\end{center}

\textbf{Abstract}

In this study, we systematically investigate the dynamics of various hadrons namely \( \pi^+ \), \( \pi^- \), \( K^+ \), \( K^- \), \( p \), \( \bar{p} \), \( \Lambda \), \( \bar{\Lambda} \), \( \Xi^- \) and \( \bar{\Xi}^+ \) produced in central Au-Au collisions. We analyze data of AGS and RHIC, which span a broad range of collision energies, ranging from \( \sqrt{s_{NN}}\) = 1.9 to 200 GeV. To analyze the transverse momentum (\( p_T \)) and transverse mass (\( m_T \)) distributions, we employ a two-component standard distribution function, achieving a very good representation of the experimental data across these energy regimes. We extract key thermodynamic parameters, including the effective temperature \( T \), the mean transverse momentum \( \langle p_T \rangle \), and the initial temperature \( T_i \), and analyze their dependence on the values of collision energy and particle mass. Our findings reveal a distinct transition behaviour around \( \sqrt{s_{NN}} = 19.6 \) GeV. Below \( \sqrt{s_{NN}} = 19.6 \) GeV, the values of \( T \), \( \langle p_T \rangle \), and \( T_i \) increase monotonically for all hadrons due to higher energy transfer into the system. Above this energy threshold, these extracted parameters plateau, suggesting that the additional energy is utilized as latent heat for phase transition rather than increasing the system's temperature. These observations delineate two distinct regions: a hadron-dominated region at lower energies and a parton-dominated region at higher energies, each potentially indicative of different phases of matter, with the latter possibly signalling the onset of a Quark-Gluon Plasma (QGP). The study thus provides critical insights into the complex interplay of thermodynamics, phase transitions, and particle interactions in high-energy Au-Au collisions.


\vskip1.0cm

\begin{multicols}{2}

{\section{Introduction}} 
While it is hard to directly probe conditions that existed 13.8 billion years ago at the dawn of our universe, contemporary physics has developed ingenious methods to recreate similar environments. One such avenue is the formation of a Quark-Gluon Plasma (QGP), a high-temperature, high-density state of matter achieved by colliding heavy ions at facilities like the Large Hadron Collider (LHC) and the Relativistic Heavy Ion Collider (RHIC) \cite{1, 2, 3}. Upon collision, the nucleons within the interacting nuclei disintegrate, resulting in a fireball of quarks and gluons. This phenomenon signifies the phase transition from hadronic matter to a state of QGP. Interestingly, during this phase transition, the temperature of the Quantum Chromodynamics (QCD) matter exhibits unique behaviour. Despite increases in collision energy or centrality, the temperature remains remarkably stable above a certain energy threshold, typically a few GeV \cite{4, 5}. This stability is attributed to the system's absorption of excess energy as latent heat, akin to how the temperature of water remains constant at its boiling point while absorbing additional heat to break molecular bonds.
The Alternating Gradient Synchrotron (AGS) and RHIC have conducted experiments over a diverse energy spectrum, ranging from approximately 2A GeV (2 GeV per nucleon) to 200 GeV, specifically in Au-Au collisions at varying levels of centrality. This research undertakes a rigorous analysis of data from AGS and RHIC to identify the collision energies that are most likely to induce a phase transition to a deconfined state of QCD matter. 

Following the formation of QGP, the system undergoes a rapid expansion due to the substantial pressure gradients present in the collision zone. This expansion leads to cooling and, ultimately, hadronization. The process from collision initiation to hadronization involves several key stages, each characterized by specific temperatures.
Initially, when the nuclei commence collision and the QGP starts to form, the relevant temperature is the initial temperature (\( T_i \)). Subsequently, the system reaches the chemical freeze-out stage, where inelastic scattering ceases. The temperature at this juncture is known as the chemical freeze-out temperature (\( T_{ch} \)) \cite{6, 7, 8}. Finally, the system arrives at the thermal or kinetic freeze-out stage, characterized by the thermal or kinetic freeze-out temperature (\( T_0 \)) \cite{9,10}. At this point, the particles become sufficiently dispersed, rendering even elastic interactions among them negligible. 
In addition to these temperatures, the concept of an effective temperature (\( T \)) is introduced, which encompasses both thermal and flow effects (\( \beta_T \)) in the system. It is expressed as \( T = T_0 + m_0 \beta_T \) \cite{4}, where \( m_0 \) is the mass of the observed hadron. Consistent with the system's evolutionary stages, the hierarchy of these temperatures in the literature is \( T_i > T_{ch} > T \geq T_0 \) \cite{11, 12}.
Understanding these phase transitions in the QGP is often mapped onto a phase diagram, akin to how the three phases of water are explained using pressure and temperature conditions. However, the QCD phase diagram, which outlines the transitions between QGP and hadronic matter, is substantially more complex and to a large extent still a matter of speculations. It is assumed to include crossover regions and second-order phase transition boundaries, adding layers of intricacy not found in simpler systems. Despite significant progress, further research is needed to fully comprehend the QCD phase diagram. In this work, our discussion will focus primarily on first-order phase transitions.

The $p_T$ distributions of hadrons produced in collisions serve as invaluable probes for extracting details about the collision dynamics up to the kinetic freeze-out stage. These distributions shed light on the properties of QCD matter, hadronic matter, and the phase transitions between them. To extract key parameters like \( T_i \), \( T \), \( T_0 \), mean $p_T$ (\( \langle p_T \rangle \)), and the transverse flow velocity (\( \beta_T \)), various statistical, hydrodynamic, and hybrid models are employed. These models include, but are not limited to, the Blast Wave Model with Boltzmann-Gibbs Statistics (BGBW model) \cite{13, 14}, Blast Wave Model with Tsallis Statistics (TBW model) \cite{15, 16}, Tsallis-Pareto type function \cite{4}, Hagedorn Model with embedded radial flow velocity \cite{17}, Hagedorn thermal model \cite{18, 11}, and the Standard Distribution function \cite{11}. Each model relies on its own set of free parameters and has a specific domain of applicability.

The manuscript is organized as follows: Section 2 elaborates on the formalism of the two-component standard distribution function that we employ in this study. Section 3 is devoted to discussing our results and the trends in the extracted parameters. The final section offers the conclusion and summarizes the key findings of the analysis.
\\
{\section{The method and formalism}} 
The standard distribution function is rooted in classical and quantum statistics, incorporating Boltzmann-Gibbs, Bose-Einstein, and Fermi-Dirac statistics. The probability density function of the standard distribution, dependent on \( p_T \), is expressed as \cite{11}

\begin{align}
f_{p_T}(p_T, T)=&\frac{1}{N}\frac{\mathrm{d}N}{\mathrm{d}p_\mathrm{T}}=C p_T \sqrt{p_T^2 + m_0^2} \nonumber \\
& \times \bigg[\exp \bigg(\frac{\sqrt{p_T^2 + m_0^2} - \mu}{T}\bigg) + S\bigg]^{-1}.
\end{align}
Or
\begin{align}
f_{p_T}(p_T, T)=\frac{1}{N}\frac{\mathrm{d}N}{\mathrm{d}p_\mathrm{T}}=C p_T \sqrt{p_T^2 + m_0^2}\int_{min}^{max}\cosh{y}\times\nonumber \\
\bigg[\exp \bigg(\frac{\sqrt{p_T^2 + m_0^2} \cosh{y}-\mu}{T}\bigg) + S\bigg]^{-1} dy.
\end{align}
Here $y$ is the rapidity, \( m_0 \) denotes the mass of the produced hadron, \( T \) is the effective temperature, \( S \) is the index of the standard distribution ($1$ for fermions and $-1$ for bosons), $C$ is the normalization constant. Eq. (2) represents the specific form of the standard distribution which is consistent with the ideal gas model \cite{18a}. \( \mu \) represents the baryon chemical potential \cite{19}. The chemical potential \( \mu \) for a particular particle species is given by \cite{20, 21}

\begin{align}
\mu_i = -\frac{1}{2} T_{ch} \ln({k_i}),
\end{align}
Where \( T_{ch} \) is the chemical freeze-out temperature, \( k \) refers to the ratio of negative to positive hadrons and $i$ in the subscript represents the particle type. The chemical freeze-out temperature \( T_{ch} \) is determined by \cite{22, 23}

\begin{align}
T_{ch} = \frac{T_{lim}}{1 + \exp[{2.60 -  \frac{\ln{\sqrt{s_{NN}}}}{0.45}}]}.
\end{align}

Here, \( T_{lim} \) represents the limiting or saturation temperature, often considered to be 0.158 GeV \cite{24}, while \( \sqrt{s_{NN}} \) is the collision energy per nucleon pair in units of GeV in the centre of mass frame.

The single-component standard distribution does not cover the high $p_T$ region, typically $p_T$ $>$ 1.5 GeV/c, therefore, two, three or even multi-component standard distribution, given below, is used \cite{11, 25} 
\begin{align}
f_{p_T}(p_T, T)=\frac{1}{N}\frac{\mathrm{d}N}{\mathrm{d}p_\mathrm{T}}=\sum_{i=1}^{n} K_i C_i p_T \sqrt{p_T^2 + m_0^2}\nonumber \\
\int_{min}^{max}\cosh{y}\times \bigg[\exp \bigg(\frac{\sqrt{p_T^2 + m_0^2} \cosh{y}-\mu}{T_i}\bigg) + S\bigg]^{-1} dy.
\end{align}
Here $K$ is the contribution fraction of the one component to the other component, while the subscript $i$ represents the $i^{th}$ number of component of the model.

From $\sqrt{s_{NN}}$ = 1.9 -- 62.4 GeV we take into account the finite values of $\mu$ in our model but beyond $\sqrt{s_{NN}}$ = 62.4 GeV, due to the almost balanced production of the particles and their anti-particles, the very small value of $\mu$ is neglected, and Eq. (5) can be written as,
\begin{align}
f_{p_T}(p_T,T)|_{\mu = 0}=\frac{1}{N}\frac{\mathrm{d}N}{\mathrm{d}p_\mathrm{T}}=\sum_{i=1}^{n} K_i C_i p_T \sqrt{p_T^2 + m_0^2}\nonumber \\
\int_{min}^{max}\cosh{y}\times \bigg[\exp \bigg(\frac{\sqrt{p_T^2 + m_0^2} \cosh{y}}{T_i}\bigg) + S\bigg]^{-1} dy.
\end{align}
In this work, we used a two-component standard distribution (Eq. (5) or Eq. (6)). The first component contributes a large fraction to the resonance decays in the very low $p_T$ region, while the second component has the contribution from hard scattering. It should be noted that even if the hard scattering is involved, the main contributor to the particle production is still the soft excitation process.

In order to superimpose the two components, the superposition principle has been used which is given by Eq. (7) \cite{25, 25a, 25b}. To formulate Eq. 7 we use the weighted average approach. The first component of the standard distribution function can only satisfy the $p_T$ spectra $<$ 1.5 GeV/c, while the second component of the same function satisfies the $p_T$ spectra $>$ 1.5 GeV/c. The first component is characterized by $T_1$ and the second one is by $T_2$. To get the resultant temperature ($T$) of the whole $p_T$ range, a weighting factor $K$ is introduced such that, 0 $\leq K \leq$ 1 representing the fraction of $T_1$ in $T$. As the maximum contribution fraction is 1 and the first component contributes $K$, therefore, it’s natural that the contribution fraction of $T_2$ is $1-K$ and hence one can deduce the following equation for $T$  
\begin{align}
T = K T_1 + (1-K) T_2.
\end{align}
This equation represents a linear interpolation between $T_1$ and $T_2$, where $K$ acts as the interpolation parameter.

Regardless of the form of particle momentum distribution, the $p_T$ dependent probability density function is given as follows,
\begin{align}
f(p_T) = \frac{1}{N}\frac{dN}{dp_T}.
\end{align}
Which is naturally normalized to unity.
\begin{align}
\int_{0}^{\infty}f(p_T)\ dp_T = 1.
\end{align}
The mean transverse momentum, $\langle p_T\rangle$, can be obtained directly from the fit function by using the probability density function in the following form,
\begin{align}
\langle p_T \rangle = \frac{\int_{0}^{\infty}p_T\ f(p_T)\ dp_T}{\int_{0}^{\infty}f(p_T)\ dp_T}.
\end{align}
Incorporating Eq. (9) we get,
\begin{align}
\langle p_T\rangle = \int_{0}^{\infty}p_T\ f(p_T)\ dp_T.
\end{align}
Similarly, the root mean square $p_T$ is given as, 
\begin{align}
\sqrt{\langle {p_T}^2 \rangle} = \sqrt{\int_{0}^{\infty}{p_T}^2\ f(p_T)\ dp_T}.
\end{align}

The temperature $T_i$ of the emission sources can be calculated by using the String Percolation model \cite{26}  
\begin{align}
T_i = \frac{p_{T (rms)}}{\sqrt{2}},
\end{align}
where $p_{T (rms)}$ = $\sqrt{\langle p_T^2\rangle}$ is the root mean square value of $p_T$.

{\section{Results and discussion}}
The double-differential \( m_T \) and \( p_T \) distributions of various produced particles in central Au-Au collisions, spanning a collision energy range of \( \sqrt{s_{NN}} = 1.9 - 200 \) GeV, are depicted in Fig. 1(a) through Fig. 1(j). The experimental data for identified particles (\( \pi^+ \), \( \pi^- \), \( K^+ \), \( K^- \), \( p \) and \( \bar{p} \)) have been obtained from \cite{27, 28, 29, 5, 30, 31, 32, 33} and for strange particles (\( \Lambda \), \( \bar{\Lambda} \), \( \Xi^- \) and \( \bar{\Xi}^+ \)) from \cite{34, 35, 36, 37}. In these figures, distinct colours and geometric shapes are employed to represent different collision energies. Filled shapes signify particles, while hollow shapes denote anti-particles. Solid lines represent the individual fits to the data, obtained using Eq. (5) or (6). Each plot also features a Data/Fit ratio panel at the bottom to indicate the quality of the individual fit. The model yields good agreement with the experimental data. To enhance readability and distinguish between data points and fit lines for varying energies, scaling factors are applied. These factors, along with other extracted parameters and the \( \chi^2/NDF \) values, are compiled in Table 1, in the Appendix. It is noteworthy that $C$ is the normalization constant which equalizes the integral of Eq. 1 to unity, i.e., it arises from $\int_{0}^{\infty} f_{p_T}(p_T, T)\ dp_T = 1$. It is different from the normalization constant $N_0$ that appeared in Table 1 in the Appendix which is equivalent to the multiplicity parameter. $N_0$ is the normalization constant that arises from the fitting procedure where it is used to compare the model with the experimental data. Although $C$ is different from $N_0$, the earlier is absorbable in the latter. Both $C$ and $N_0$ are used to have a precise description of the formalism and results.

In Fig. 1 (a) and (b) $|y| <$ 0.05 for $\sqrt{s}$ = 1.9 to 3.4 GeV, for 7.7 to 54.4 GeV  $|y| <$ 0.1, for 62.4 GeV $|y| <$ 0.5 and for 130 and 200 GeV $|\eta| <$ 0.35, where $\eta$ represents the pseudorapidity. In Fig. 1(c) $|y| <$ 0.25 for $\sqrt{s}$ = 1.9 to 3.4 GeV while in Fig. 1(c) and (d) at $\sqrt{s}$ = 7.7 to 54.4 GeV, $|y| <$ 0.1 and for $\sqrt{s}$ = 130 and 200 GeV, $|\eta|$ = 0.35. In Fig. 1(e) $|y| <$ 0.05 at $\sqrt{s}$ = 1.9 to 3.4 GeV, in Fig. 1(e) and (f) for $\sqrt{s}$ = 7.7 to 54.4 GeV $|y| <$ 0.1, while at $\sqrt{s}$ = 62.4 GeV $|y| <$ 0.5 and for 130 and 200 GeV $|\eta| <$ 0.35. In Fig. 1(g), (h), (i) and (j) $|y| <$ 0.5 for $\sqrt{s}$ = 7.7 to 39 GeV and $|y| <$ 1 at $\sqrt{s}$ = 62.4 GeV. In Fig. 1(g), (h), (i) and (j) $|y| <$ 0.05 at $\sqrt{s}$ = 130 GeV. $|y| <$ 1 for $\sqrt{s}$ = 200 GeV in Fig. 1(g) and (h), while in Fig. 1(i) and (j), $|y| <$ 0.75 at $\sqrt{s}$ = 200 GeV.

Although our model, as described by Eq. (6), is \( p_T \)-dependent, i.e., \( f(p_T, T) \), it can also be adapted to be \( m_T \)-dependent, i.e., \( f(m_T, T) \), for analyzing \( m_T \) spectra. This adaptability arises because the \( p_T \) and \( m_T \) spectra are nearly identical and can be used interchangeably.

\begin{figure*}[p!]
\centering
\includegraphics[width=0.3\textwidth]{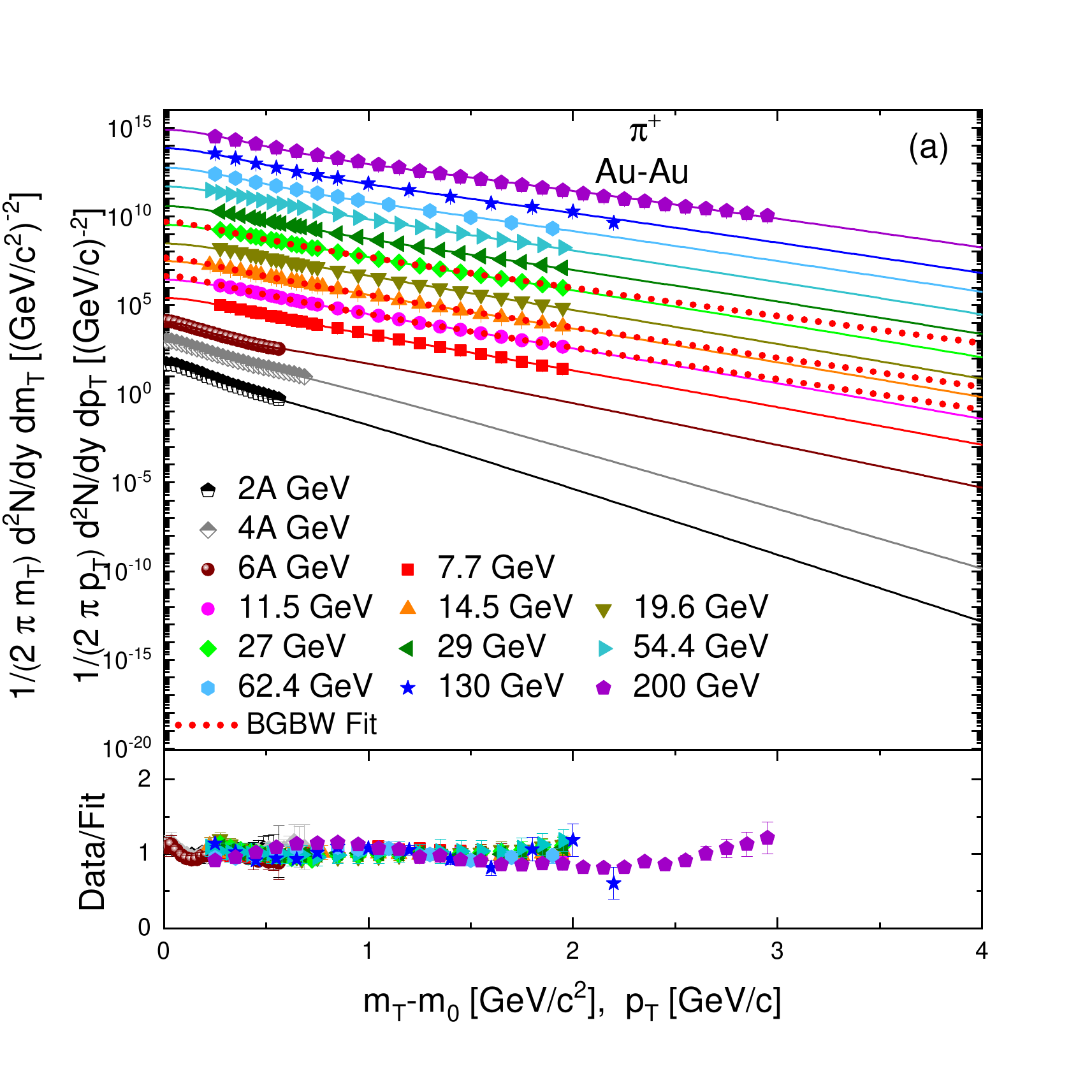}
\includegraphics[width=0.3\textwidth]{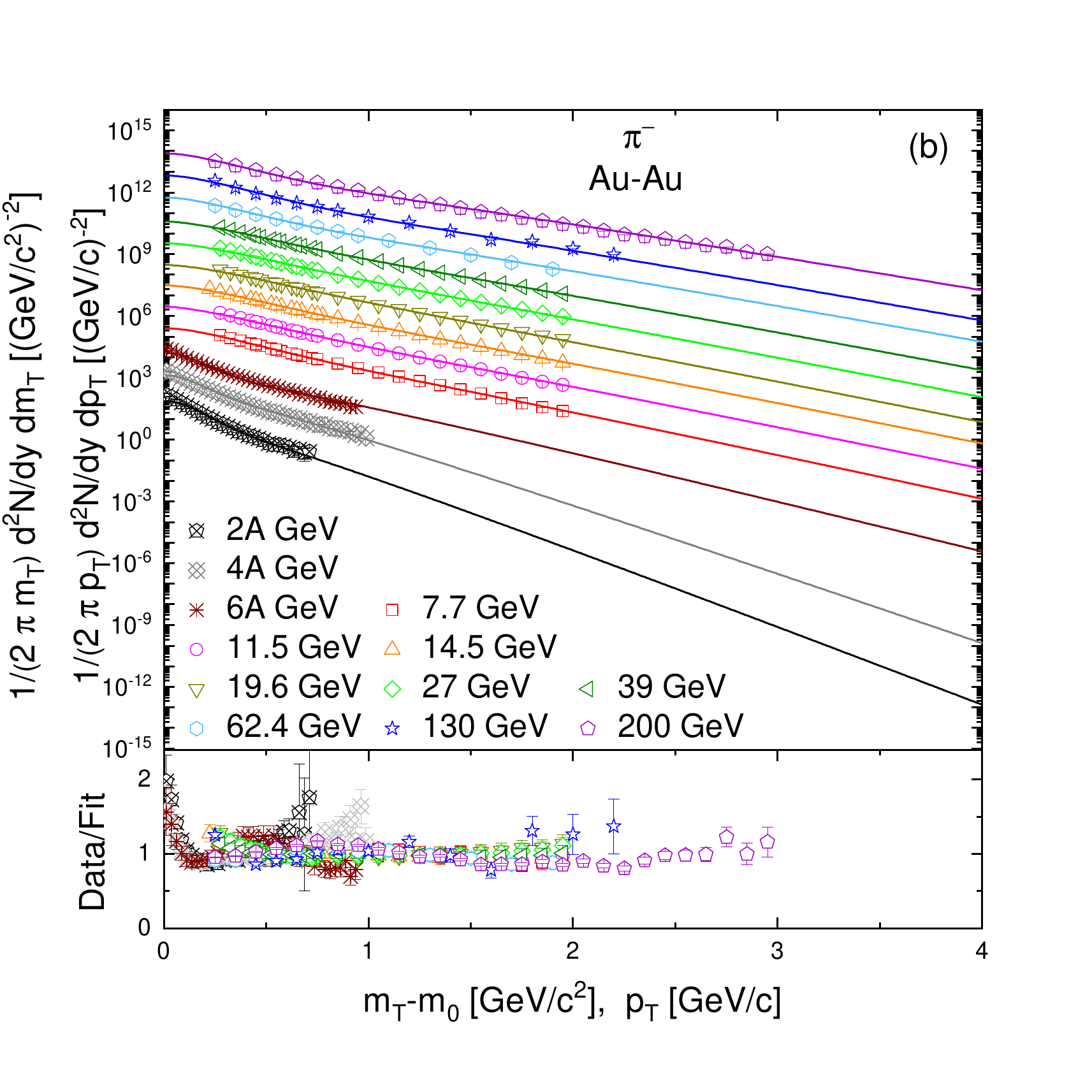}
\includegraphics[width=0.3\textwidth]{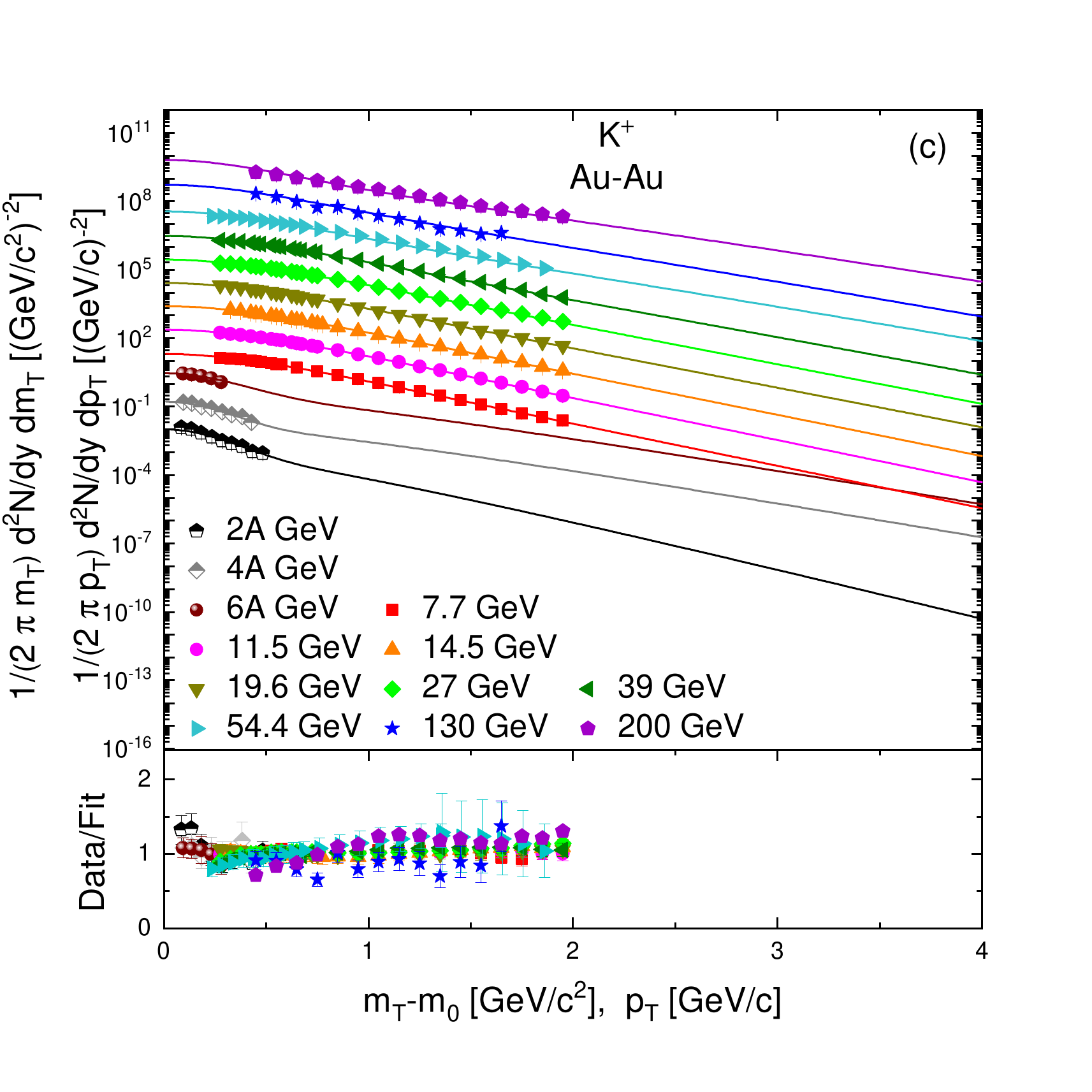}\vspace{-0.35cm}
\includegraphics[width=0.3\textwidth]{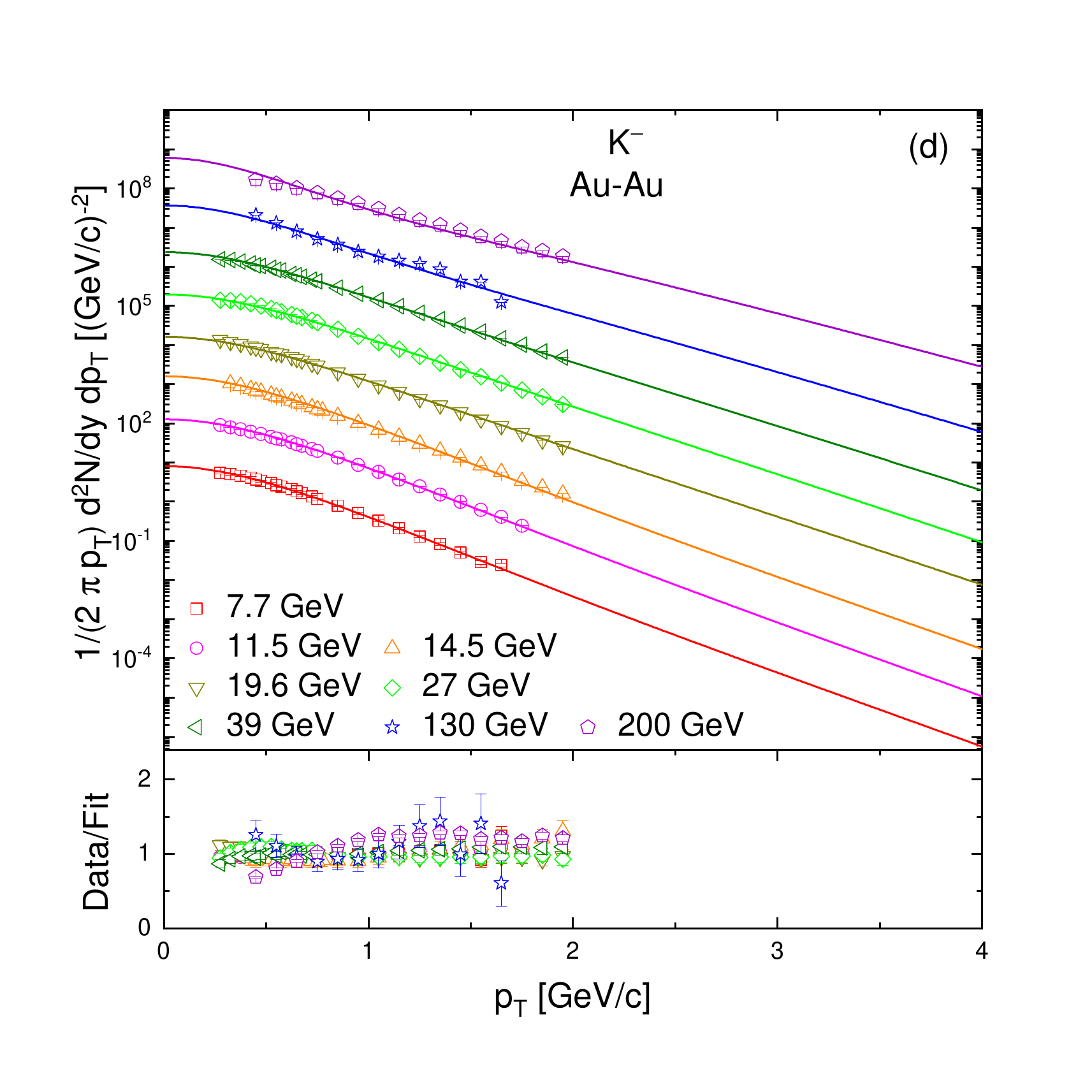}
\includegraphics[width=0.3\textwidth]{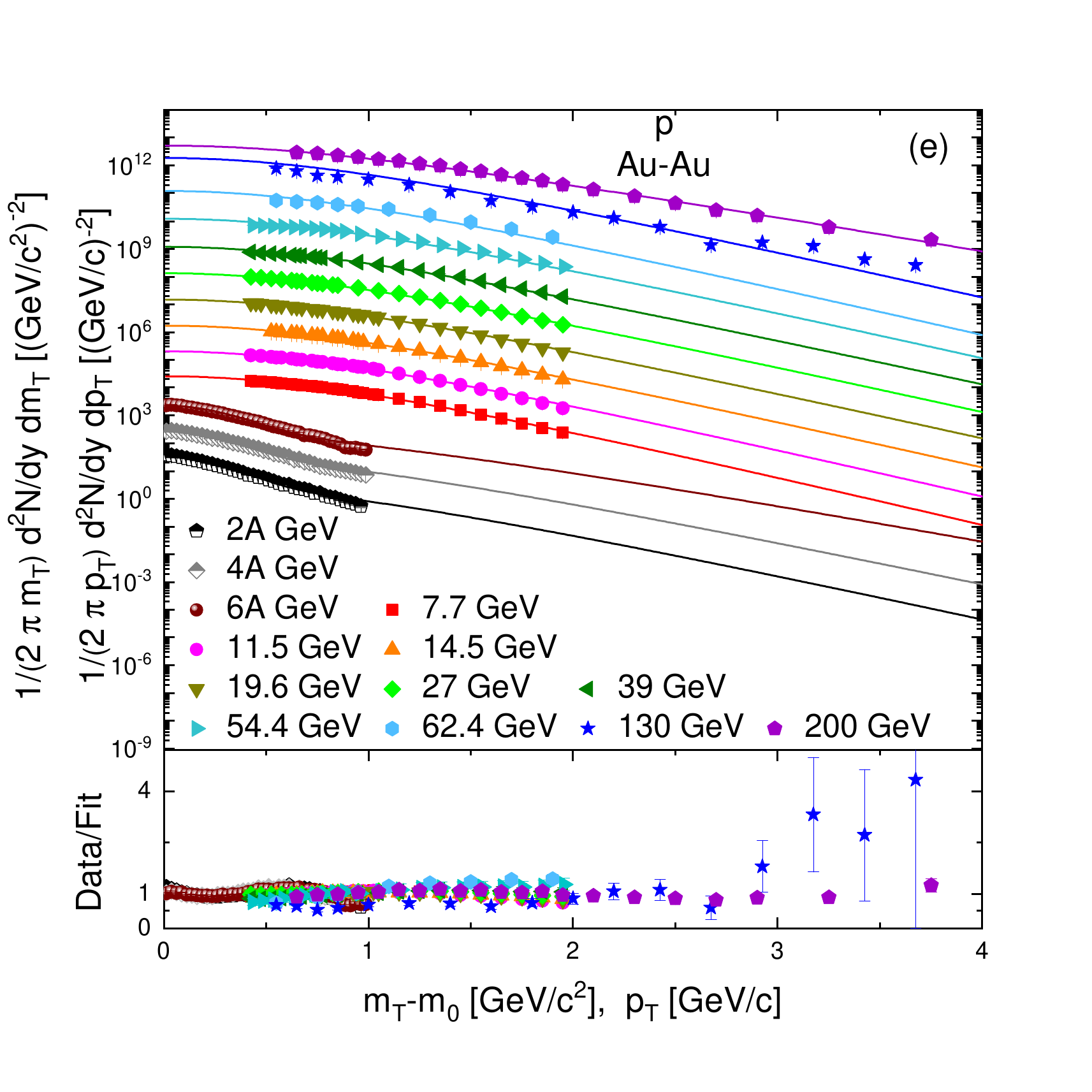}
\includegraphics[width=0.3\textwidth]{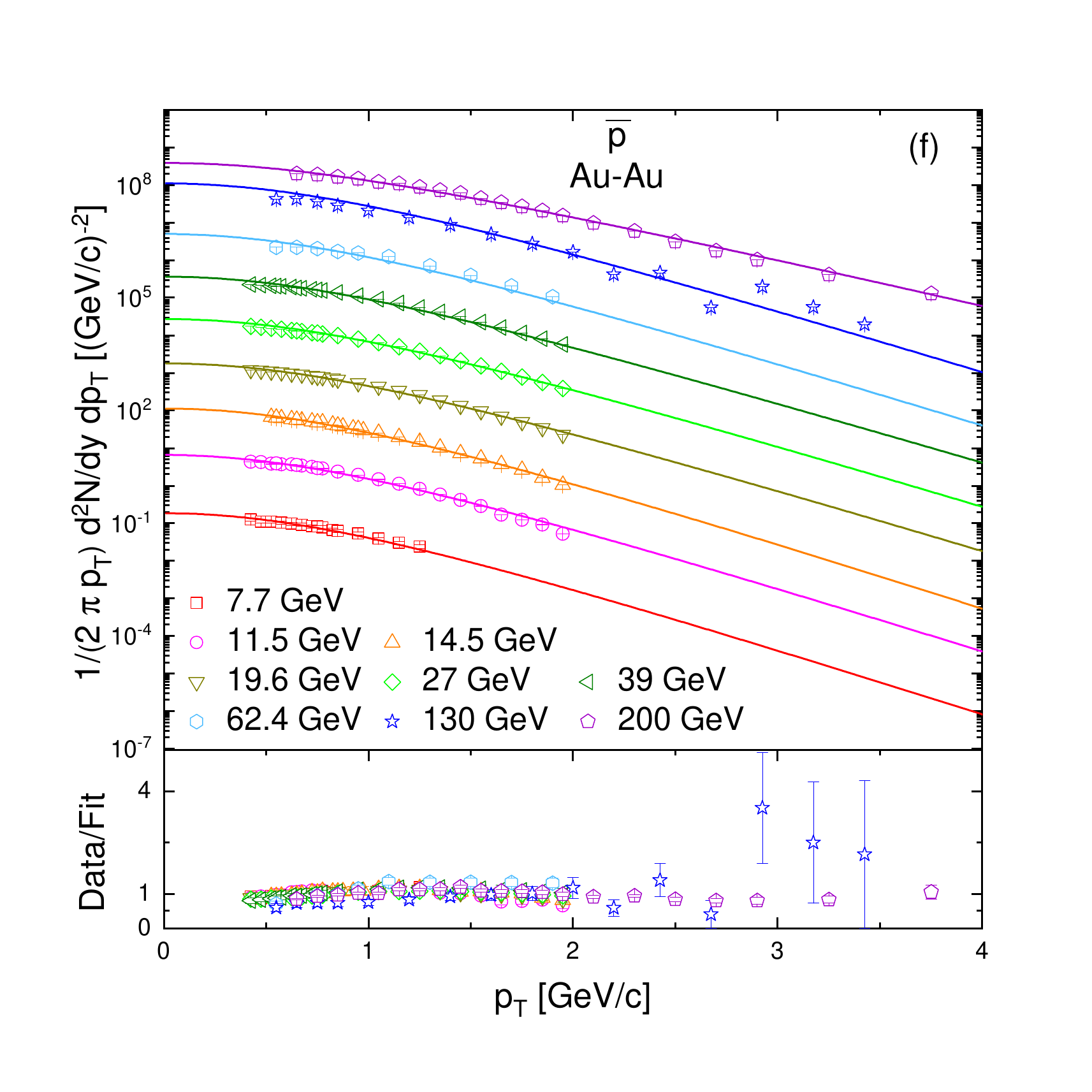}\vspace{-0.35cm}
\includegraphics[width=0.3\textwidth]{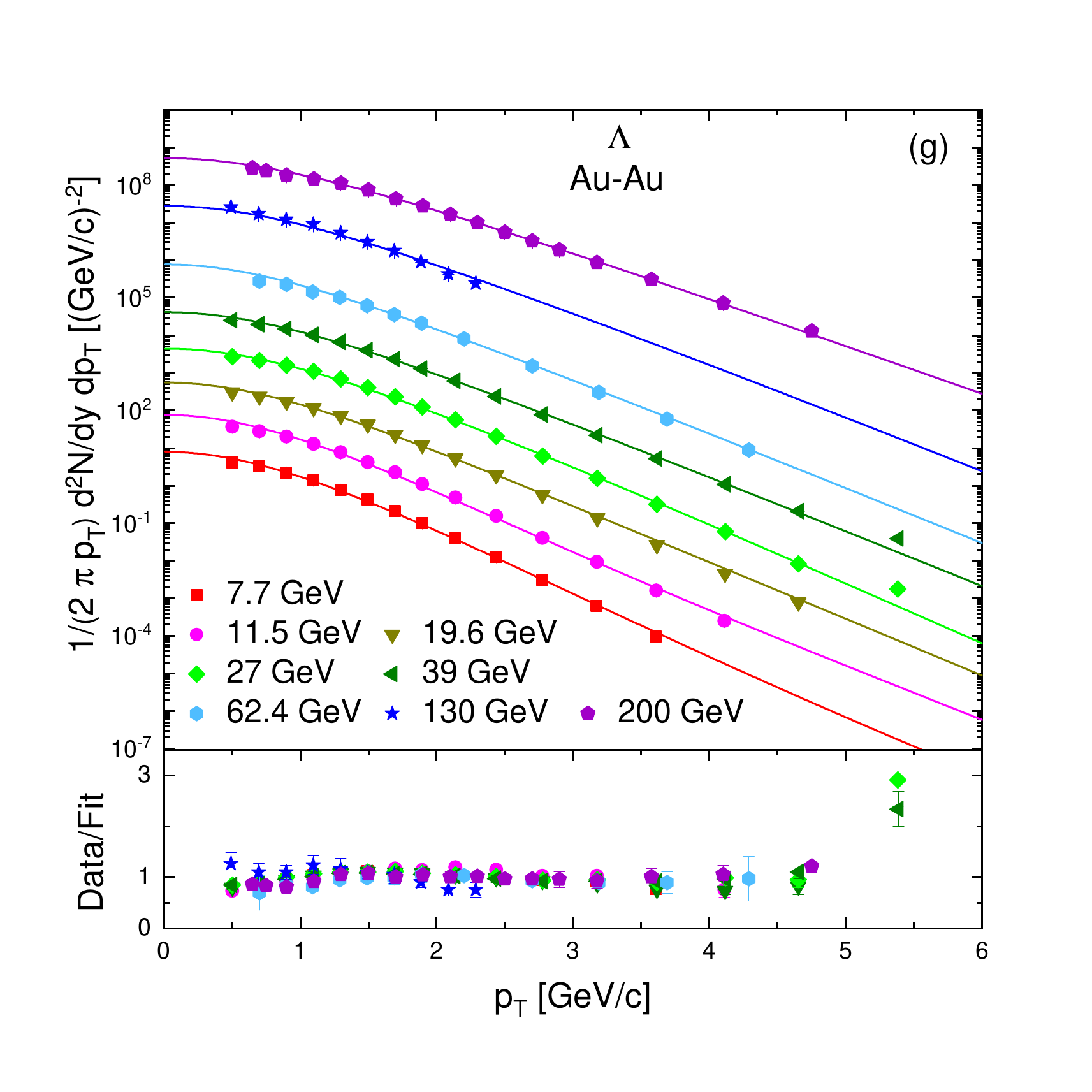}
\includegraphics[width=0.3\textwidth]{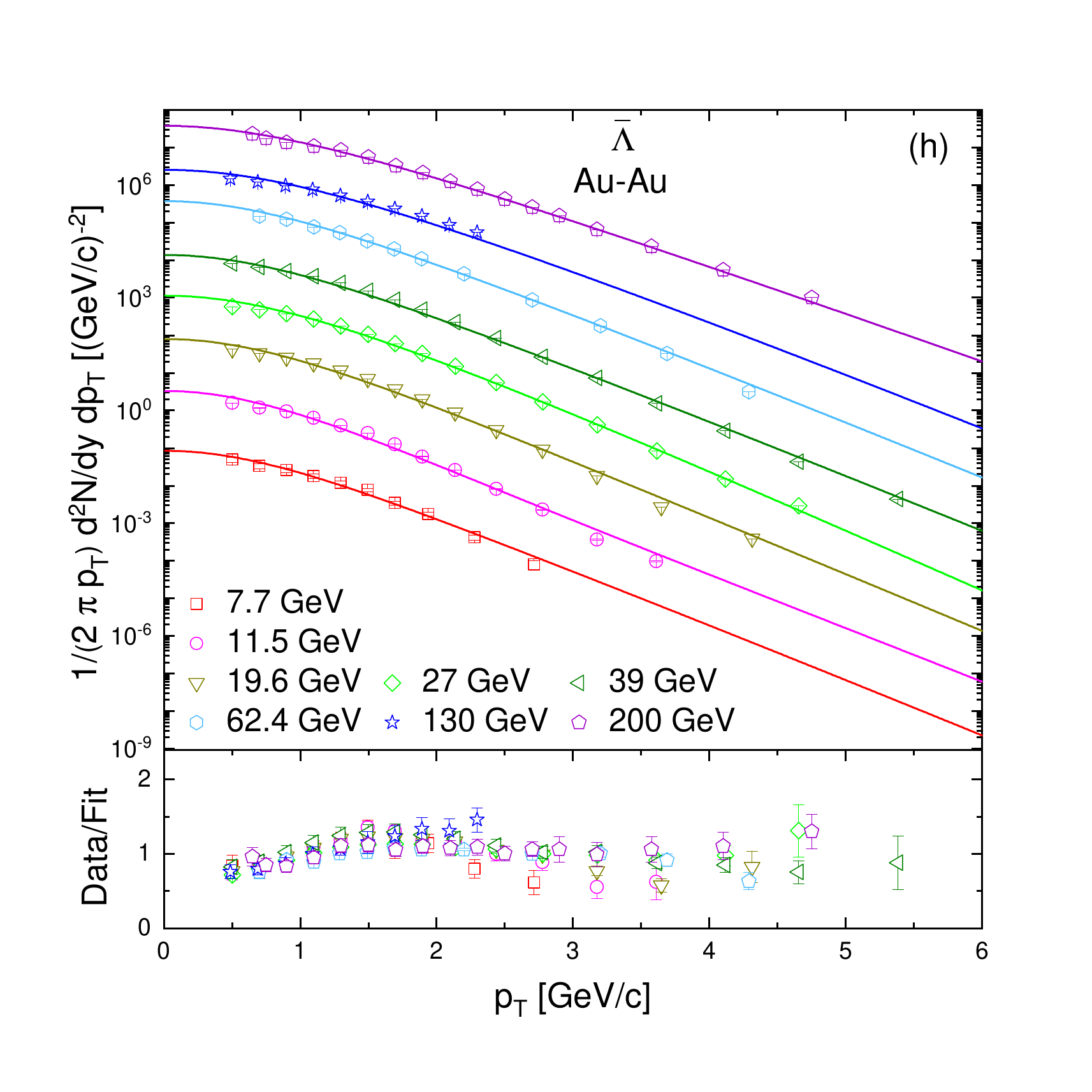}
\includegraphics[width=0.3\textwidth]{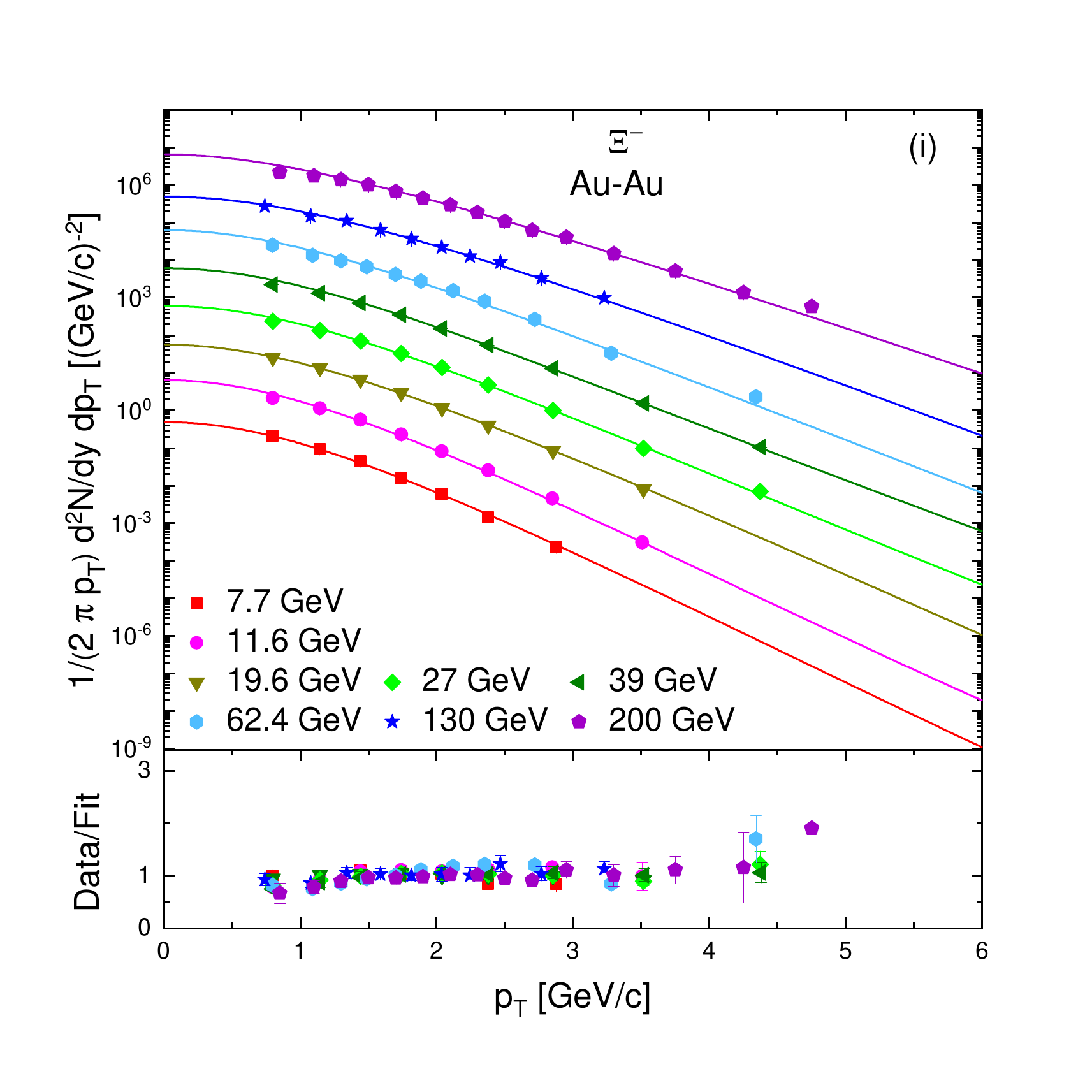}\vspace{-0.35cm}
\includegraphics[width=0.3\textwidth]{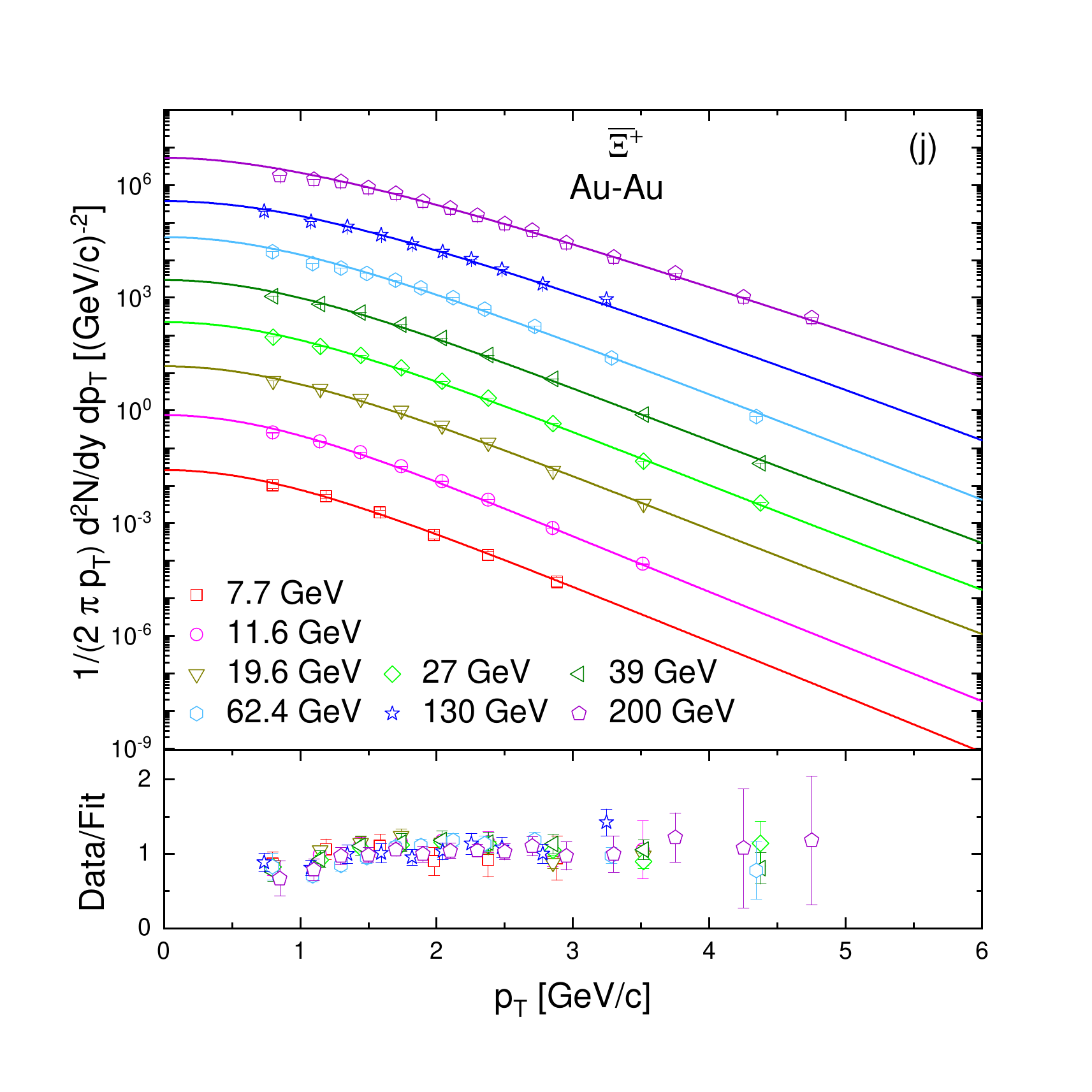}
\caption {Plots (a) - (j) display the double-differential \( m_T \) and/or \( p_T \) spectra of identified particles \cite{27, 28, 29, 5, 30, 31, 32, 33} and strange particles \cite{34, 35, 36, 37}. For AGS energies, the graphs utilize \( \frac{1}{2 \pi m_T} \frac{d^2N}{dy dm_T} \) vs \( m_T - m_0 \), while for RHIC energies, \( \frac{1}{2 \pi p_T} \frac{d^2N}{dy dp_T} \) vs \( p_T \) is used. Distinct colours and geometrical shapes indicate varying collision energies. Solid lines represent the fitting results based on the two-component standard distribution as defined in Eq. (5) or Eq. (6). The dotted lines over a few spectra in panel (a) are used for the fitted results of Boltzmann-Gibbs Blast Wave (BGBW) model. Filled symbols signify particles, and hollow symbols represent their corresponding anti-particles. To assess the quality of the fits, a Data/Fit ratio panel is included at the bottom of each plot.}
\end{figure*}
Now we explore the variations in extracted parameters as functions of both collision energy and particle mass. Fig. 2(a) depicts \( T \) in relation to the collision energy and particle mass. Different symbols and colors are employed to distinguish between particle species, with filled and hollow symbols representing particles and anti-particles, respectively. Due to the symmetry in the extracted parameters for hadrons and their corresponding anti-hadrons, the data points for the hadrons are often obscured by those for the anti-hadrons. To mitigate this, we have enlarged the point sizes for the hadrons relative to the anti-hadrons. Fig. 2(b) and 2(c) follow the same conventions as Fig. 2(a) but display \( \langle p_T \rangle \) and \( T_i \), respectively. It is evident that \( T \), \( \langle p_T \rangle \), and \( T_i \) exhibit similar trends as functions of both energy and particle mass.

In Fig. 2(a), we observe that \( T \) sharply increases from lower energies up to \( \sqrt{s_{NN}} < 19.6 \) GeV, and gives a plateau thereafter up to the maximum energy of 200 GeV. This behaviour can be explained by considering two distinct energy regimes. In the first regime (\( \sqrt{s_{NN}} < 19.6 \) GeV), the system is baryon-dominated and lacks the energy required for a phase transition. As the collision energy increases, the system transitions from being baryon-dominated to parton-dominated, indicative of a phase transition. Specifically, \( T \) rises from lower energies up to 19.6 GeV due to increased energy deposition in the system and hence the system attains a higher degree of excitation. $T$ remains constant beyond 19.6 GeV, and this plateau suggests that additional energy serves as latent heat because any energy supplied to the system beyond 19.6 GeV is totally used to melt the bonds among partons within hadrons. It is not used to rise $T$, \( \langle p_T \rangle \) and \( T_i \) and hence facilitating the phase transition to QGP from hadronic matter. The system begins its phase transition in a portion of the volume at 19.6 GeV, and as a result of the phase transition in progressively larger volumes between 19.6 and 200 GeV, it shifts from being hadron-dominated to parton-dominated. The partial phase transition begins at 19.6 GeV, and the full phase transition begins at 200 GeV. The 19.6–200 GeV energy range is the critical energy range.

It is important to note that the values of \( T \) may appear larger which may be due to the model used. The employed functions are regarded as thermometric scales. Since different functions employ various thermometric scales, their values—whether high or low—are comparatively different from one another. We need a way to bring disparate thermometric scales in subatomic physics together, similar to the thermometric scale used in thermal and statistical physics. However, structuring the approach is outside the scope of the present work.

Different studies using various models have reported conflicting results concerning the onset of phase transitions, making it an open question in the community. For instance, some works report phase transition at energies as low as 7.7 GeV or as high as 14.5 GeV \cite{41, 5, 42}. Additionally, there is an ongoing debate about the behaviour of \( T \) at energies beyond 39 GeV.
Our current study aims to resolve some of these discrepancies by analyzing a diverse set of particle species—identified particles, strange particles and multi-strange particles. We find that \( T \) is symmetric across all particles and their corresponding anti-particles, in contrast to our previous work where light nuclei showed a separate freeze-out \cite{42, 43}. Moreover, \( T \) appears to be mass-dependent: heavier particles exhibit a higher \( T \) compared to lighter particles. This finding confirms the multiple-freeze-out scenario and aligns with our earlier research \cite{44, 45}. This finding suggests that all particles have different freezeout mechanisms and thermalization processes.

By weighting the yields of various particles, one can determine the global (simultaneous) fit from the average of the parameters. Although individual fitting to transverse momentum spectra is frequently used, there is indeed no particle dependence when performing global fitting. However, by performing individual fits, researchers can extract specific information about the properties and ascertain the features of the underlying particle production mechanisms. This fitting method makes it possible to comprehend the physics of collision processes and the characteristics of the particles that are created.  Additionally, individual fitting permits researchers to independently examine different aspects of the collision, such as distinct particle species or collision centrality classes. This degree of detail can help in separating the different contributions to the observed spectra and offer a more comprehensive understanding of the physics at play. It is also worth mentioning that, although the actual values of $T$ would change by performing the global fit, however, the trend of the parameters with the centre of mass energy and centrality does not change, as reported in our previous work \cite{42}. Moreover, in the present study, we analyze a vast variety of particle species, therefore, the fit quality in simultaneous fit may be poor.

Figs. 2(b) and 2(c) display trends for \( \langle p_T \rangle \) and \( T_i \), respectively, that closely mirror the behavior observed for \( T \). \( \langle p_T \rangle \) increases with the collision energy which shows that large momentum has been transferred to the system which results in further multiple scatterings. Above 19.6 GeV, \( \langle p_T \rangle \) remains constant like $T$ indicating the phase transition. Similarly, $T_i$ increases from lower energies up to 19.6 GeV which shows that the initial energy density increases from lower energies up to 19.6 GeV, however, it renders the phase transition at 19.6 GeV. The similar behavior of $T$, \( \langle p_T \rangle \) and $T_i$ shows that \( \langle p_T \rangle \) and $T$ are the reflections of $T_i$ and it agrees with the system evolution process. It should be noted that like $T$, $T_i$ and \( \langle p_T \rangle \) are observed to be mass dependent, and they are larger for massive particles. However, the situation is different from our previous work \cite{46}, where $T_i$ and \( \langle p_T \rangle \) are mass dependent while $T_0$ is dependent on the cross-section interaction of the particle.

\begin{figure*}[p!]
\centering
\includegraphics[width=0.49\textwidth]{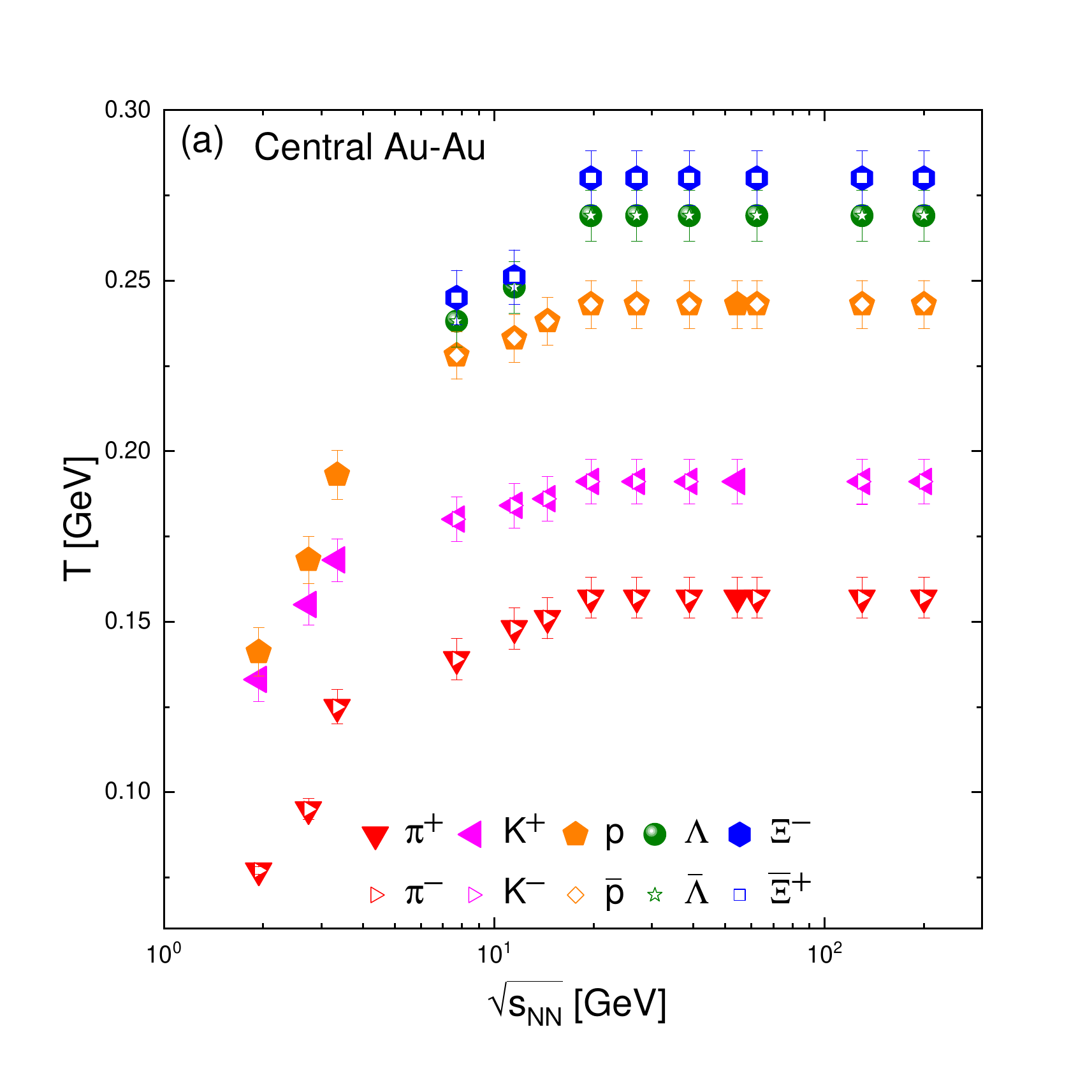}\vspace{-0.5cm}
\includegraphics[width=0.49\textwidth]{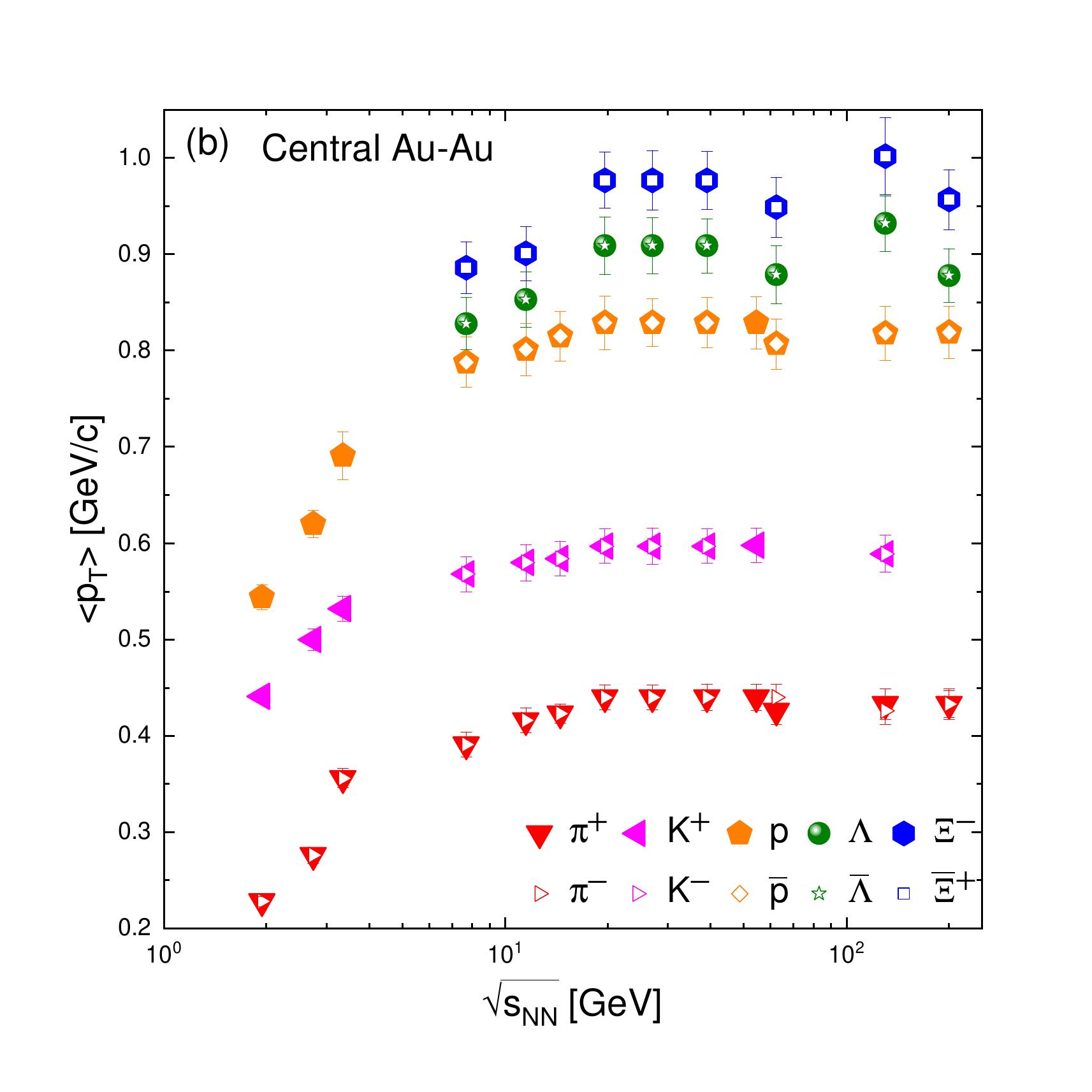}
\includegraphics[width=0.49\textwidth]{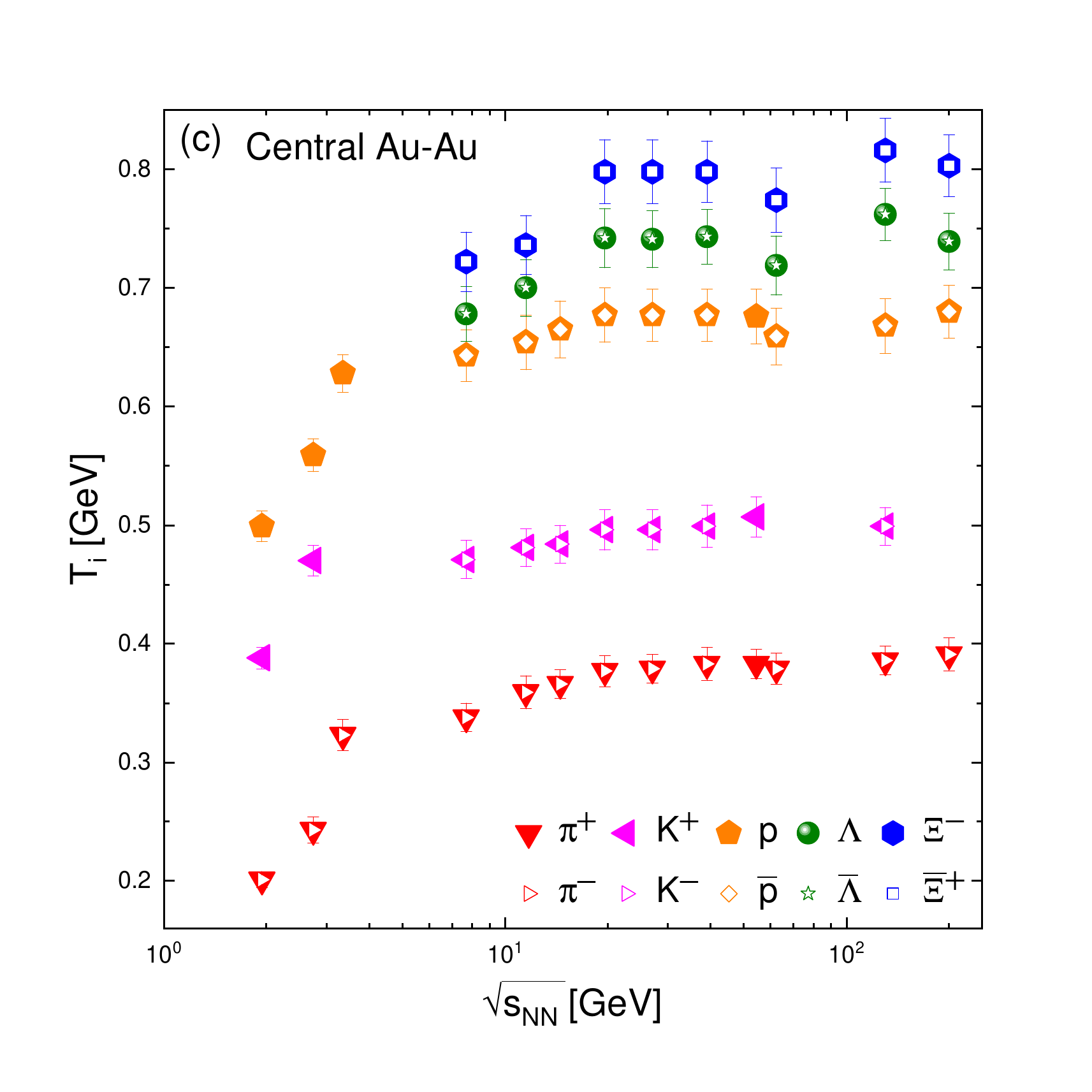}\vspace{-0.5cm}
\caption {In this figure, three key parameters are plotted as functions of collision energy: (a) \( T \), (b) \( \langle p_T \rangle \), and (c) \( T_i \). Various symbols and colors are used to denote different types of observed particles. Filled symbols represent particles, while hollow symbols are used for their corresponding anti-particles. This graphical representation allows for a detailed examination of how these crucial thermodynamic and kinematic variables evolve with collision energy and particle type.}
\label{fig2}
\end{figure*}

\begin{figure*}
\centering
\includegraphics[width=0.49\textwidth]{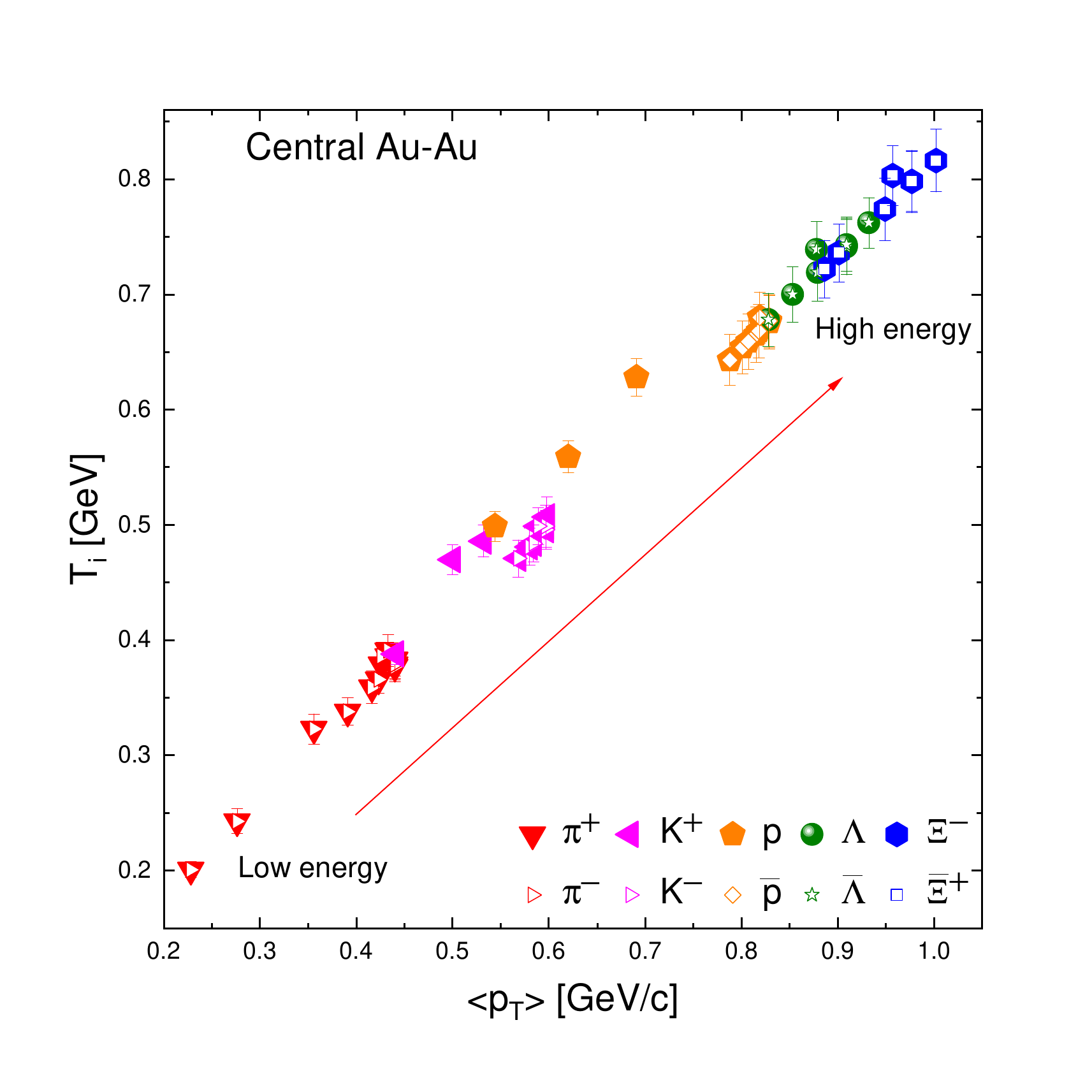}\vspace{-0.5cm}
\caption { Correlation of $T_i$ and $\langle p_T \rangle$.}
\label{fig3}
\end{figure*}

In Fig. 3, the positive correlation between \( T_i \) and \( \langle p_T \rangle \) is evident, with \( T_i \) plotted on the y-axis and \( \langle p_T \rangle \) on the x-axis. This trend is corroborated by the data in Table 1, which also demonstrates a positive correlation between \( T \), \( T_i \), and \( \langle p_T \rangle \). Interestingly, Fig. 3 reveals a noteworthy phenomenon concerning particle freezeout times. Specifically, lighter particles at high energies exhibit thermodynamic behaviour akin to that of heavier particles at lower energies. For instance, pions at higher energies align with kaons at lower energies, and similarly, \( K \) at high energies aligns with \( p \) at low energies. This suggests a form of thermodynamic equivalence between particle types under varying energy conditions.

$T$ represents the effective temperature throughout the manuscript, as it contains both thermal and flow effects and is generally given as \cite{4, 47}
\begin{align}
 T=T_0+m_0\ \beta_T.     
\end{align} 								
 Where $T_0$ is the kinetic freeze-out temperature, $m_0$ is the rest mass of the particle and $\beta_T$ is the transverse flow velocity of the expanding source. In the literature \cite{4, 48} it has been found that both $T_0$ and $\beta_T$ remain constant as phase transition starts from hadronic to QCD matter. This implies that $T$ also remains invariant during such a phase transition in agreement with Eq. (14). To verify Eq. (14), we use the conventional Blast Wave model \cite{13} to the $p_T$ spectra of $\pi^+$ at three different collision energies. At $\sqrt{s_{NN}}$ = 11.5 GeV, we obtain $T_0$ = 0.066 GeV, and $\beta_T$ = 0.590. Inserting these values of $T_0$ and $\beta_T$ in Eq. (14), the value of $T$ turns out to be 0.145 GeV, that is exactly what we obtained for $T$ by Eq. (5). At $\sqrt{s_{NN}}$ = 14.5 GeV, $T_0$ = 0.068 GeV, and $\beta_T$ = 0.595. Using these values of $T_0$ and $\beta_T$ in Eq. (14), the value of $T$ turns out to be 0.151 GeV which is exactly the value of $T$ we have extracted from Eq. (5). Similarly, at $\sqrt{s_{NN}}$ = 27 GeV the extracted values of $T_0$ and $\beta_T$ are 0.073 GeV and 0.596, respectively. Using again these values in Eq. (14) we obtain $T$ = 0.157, the same as the one we have extracted from Eq. (5).

 Unlike temperature, the volume of the system is subjected to changes during the phase transition. For example, during the phase change of water (liquid) into steam (gas) the temperature of water at its boiling point (100 degrees Celsius at standard conditions of pressure) remains invariant but due to the conversion of water into steam, its volume abruptly increases. If the changes in volume are continuous or gradual the transition is a cross-over phase transition but if the changes are discontinuous or abrupt the phase transition is a first-order phase transition.
The normalization constant or multiplicity parameter, $N_0$, is directly related to the volume, V, of the produced system by the equation \cite{17}, $V = (2 \pi)^3 N_0/g$, where $g$ is the degeneracy factor which is 1 for pion and kaon and 2 for proton. The discontinuous changes in $N_0$ during the phase transition suggest the phase transition to be discontinuous and more likely of the first order. Note that one can apply the derivative analysis to the data set of multiplicity parameters to check its discontinuity.

Believing the phase transition to be a cross-over above the collision energies of 10A GeV is just a theoretical prediction with several serious implications. For instance, in the cross-over region (a region where two states of matter gradually convert into each other by continuous changes in the physical properties like temperature etc.) both hadron gas and free quarks can co-exist. This means that we can have full quark confinement (only hadrons with no free quarks) only at absolute zero temperature, might not be achievable, and at any temperature above absolute zero \cite{49}, no matter how small is that there will be liberated quarks along with hadrons. If this is the case, why so far a free quark has not been experimentally observed at ordinary temperature? This being the case, we believe that, in the physical world, the transition from the confined to the deconfined phase is a discontinuous phase transition, most likely of first order. That is why we believe that the constancy of temperature and other parameters with collision energy is one of the possible signs of the first-order phase transition.

We would like to emphasize that while the model, when used independently, may not be sufficient to yield information about the deconfinement phase transition from hadronic matter to QGP, the excitation function of the extracted parameter is anticipated to exhibit specific tendencies. These tendencies encompass but are not confined to, peak and valley structures, rapid and gradual variations, positive and negative changes, and constancy in variations, among others. These patterns are intricately linked to the equation of state (EOS) of the considered matter. The alteration in EOS signifies a potential shift in the interaction mechanism, transitioning from a hadron-dominated to a parton-dominated state. It is inherent that these explanations extend beyond a singular dataset. The current model intends to outline a methodology for fitting and elucidating the data.
 
The presence of a plateau structure in the excitation function of $T$ and other parameters is anticipated to have a connection with the phase transition, although this correlation currently lacks clarity. Additional research about this matter is necessary for drawing a strong conclusion. Stating the onset of deconfinement solely based on the quality of certain fits is a tentative interpretation. A more comprehensive understanding necessitates further investigations and a comparison with other discoveries. However, the significance of the results obtained in this work can’t be denied and plays a relevant role, along with other results, in drawing a strong conclusion about the onset of phase transition.
   
\section{Conclusions}
In this comprehensive study, we analyze the \( p_T \)  (\( m_T \)) spectra of various particles produced in central Au-Au collisions. The data spans a broad spectrum of collision energies, ranging from \( \sqrt{s_{NN}} = 1.9 \) GeV up to 200 GeV. Utilizing a two-component standard distribution function, we find a good fit to the experimental data covering these different energy regimes. We extracted thermodynamic parameters, namely \( T \), \( \langle p_T \rangle \), and \( T_i \) and scrutinized for their dependence on both the collision energy and the mass of the particles involved.

A striking result is the distinct behaviour of these parameters around a collision energy of \( \sqrt{s_{NN}} = 19.6 \) GeV. Beyond this energy threshold, the parameters appear to plateau. This plateau suggests that additional energy in the system is being utilized as latent heat for a phase transition, rather than contributing to an increase in them. This behaviour delineates two separate regimes of collision energy: at $\sqrt{s_{NN}}$$<$ 19.6 GeV, the system is hadron-dominated, while as the energy increases the system becomes parton-dominated. The system starts phase transition in part of the volume at 19.6 GeV, and in the whole volume at 200 GeV, which indicates that the critical energy lies between 19.6 and 200 GeV. 

The reported values of $N_0$ (directly linked with the volume of the produced system) after the critical point of collision energy i.e., after $\sqrt{s_{NN}}$ = 19.6 GeV show the discontinuous nature suggesting the discontinuous phase transition more likely of the first order.

In addition to the energy-dependent behaviour of \( T \), \( \langle p_T \rangle \), and \( T_i \), we also observe a distinct mass-dependent trend in these parameters. Heavier particles exhibit higher \( T \), potentially indicating that they undergo decoupling from the system at an earlier stage when the system is still hot. On the other hand, lighter particles exhibit lower $T$, indicating that they decouple later from the system when the system is less thermalized comparatively. This observation could imply the existence of a multiple freeze-out scenario in central Au-Au collisions, adding another layer of complexity to our understanding of high-energy collision dynamics. Moreover, a single freeze-out has been reported for particles and their corresponding anti-particles in this study.

Additionally, we observe a positive correlation between the extracted parameters \( T \), \( \langle p_T \rangle \), and \( T_i \), which highlights the complex interplay between thermodynamics and particle interactions in high-energy collisions. Our findings provide critical insights into the thermodynamics and phase transitions occurring in these systems and address some of the existing controversies and open questions in the field. Specifically, our analysis benefits from the inclusion of a diverse array of particle species, lending further credibility to our results. Overall, this study contributes to our broader understanding of the behaviour of matter under extreme conditions. We observed similar thermodynamic properties of the lighter particles at higher energies to that of the heavier particles at lower energies.\\

\noindent \textbf{Data availability}
 The data supporting the findings of this study are fully integrated within the article and are appropriately cited as references within the text at relevant points.\\

\noindent {\bf Compliance with Ethical Standards}
The authors hereby declare that they have adhered to all ethical standards and guidelines in the preparation and presentation of the content contained in this paper.\\

\noindent {\bf Acknowledgments}
The authors gratefully acknowledge the institutional support provided by Hubei University of Automotive Technology, China, Abdul Wali Khan University Mardan (AWKUM), Pakistan, and Princess Nourah bint Abdulrahman University Researchers Supporting Project number (PNURSP2024R106), Princess Nourah bint Abdulrahman University, Riyadh, Saudi Arabia. We also extend our appreciation to the Deanship of Scientific Research at Northern Border University, Rafha, KSA, for funding this research work through the project number"NBU-FFR-2024-2439-01".
\\
\end{multicols}

\appendix
\begin{sidewaystable}
\Large\textbf{{Appendix}}\\\\\\
\small \textbf{Table 1.} The values of different parameters, extracted from the two-component standard distribution, along with the values of $\chi^2$/NDF. $C\%$ and S.F in the table are used for collision centrality percent and scaling factor, respectively. 
\scriptsize
\begin{center}
    \begin{tabular}{ccccccccccccc}
    \hline
    \hline
        \textbf{Collision} & \textbf{Particle} & \textbf{C$\%$} & \textbf{Energy [GeV]} & \textbf{$|y|$} & \textbf{S.F} & \textbf{$T_1$ [GeV]} & \textbf{$T_2$ [GeV]} & \textbf{K} & \textbf{$T$ [GeV]} & \textbf{$N_0$} & \textbf{$\chi^2$/NDF} \\ 
        \hline
        \hline
        \textbf~ & $~$ &  0-5\% & 1.9 & $<$ 0.05 & 1 & 0.111±0.002 & 0.048±0.002 & 0.454±0.021 & 0.077±0.001 & 530±21.34 & 6.0279/21 \\ 
        \textbf{} & ~ &  0-5\% & 2.7 & $<$ 0.05 & 10 & 0.124±0.003 & 0.05±0.003 & 0.604±0.022 & 0.095±0.003 & 1320±50.43 & 9.7103/26 \\ 
        \textbf{} & ~ & 0-5\% & 3.4 & $<$ 0.05 & $10^2$ & 0.171±0.005 & 0.054±0.004 & 0.604±0.021 & 0.125±0.004 & 2090±80.35 & 17.0664/21 \\ 
        \textbf{} & ~ & 0-5\% & 7.7 & $<$ 0.1 & $10^3$ & 0.194±0.004 & 0.108±0.003 & 0.356±0.017 & 0.139±0.003 & 9260±333.36 & 11.1691/24 \\ 
        \textbf{} & ~ & 0-5\% & 11.5 & $<$ 0.1 & $10^4$ & 0.204±0.004 & 0.117±0.004 & 0.356±0.017 & 0.148±0.004 & 11260±405.36 & 4.2825/24 \\ 
        \textbf{Fig. 1 (a)} & $\pi^+$ & 0-5\% & 14.5 & $<$ 0.1 & $10^5$ & 0.209±0.005 & 0.119±0.005 & 0.354±0.016 & 0.151±0.003 & 12961±466.596 & 2.5307/26 \\ 
        \textbf{Au-Au} & ~ & 0-5\% & 19.6 & $<$ 0.1 & $10^6$ & 0.209±0.003 & 0.129±0.004 & 0.354±0.017 & 0.157±0.002 &  14060±506.16 & 19.5100/24 \\ 
        \textbf{} & ~ & 0-5\% & 27 & $<$ 0.1 & $10^7$ & 0.214±0.005 & 0.127±0.004 & 0.354±0.017 & 0.157±0.003 & 15639±563.004 & 12.1858/24 \\ 
        \textbf{} & ~ & 0-5\% & 39 & $<$ 0.1 & $10^8$ & 0.223±0.005 & 0.122±0.003 & 0.354±0.015 & 0.157±0.003 & 16960±610.56 & 3.4854/24 \\ 
        \textbf{} & ~ & 0-5\% & 54.4 & $<$ 0.1 & $10^9$ & 0.223±0.005 & 0.122±0.004 & 0.354±0.017 & 0.157±0.003 & 21650±779.4 & 3.6874/21 \\ 
        \textbf{} & ~ & 0-10\% & 62.4 & $<$ 0.5 & $10^{10}$ & 0.24±0.005 & 0.112±0.004 & 0.354±0.017 & 0.157±0.004 & 23750±855 & 6.7174/11 \\ 
        \textbf{} & ~ & 0-5\% & 130 & $|\eta|$ $<$ 0.35 & $10^{11}$ & 0.24±0.006 & 0.112±0.005 & 0.354±0.018 & 0.157±0.005 & 25450±865 & 59.3027/12 \\ 
        \textbf{} & ~ & 0-5\% & 200 & $|\eta|$ $<$ 0.35 & $10^{12}$ & 0.253±0.006 & 0.11±0.004 & 0.334±0.016 & 0.157±0.003 & 29250±1053 & 35.2011/26 \\ 
        \hline
        \textbf~ & ~ & 0-5\% & 1.9 & $<$ 0.05 & 1 & 0.111±0.002 & 0.048±0.002 & 0.454±0.021 & 0.077±0.001 & 750±31.34 & 85.0912/27 \\ 
        \textbf{} & ~ & 0-5\% & 2.7 & $<$ 0.05 & 10 & 0.124±0.003 & 0.05±0.003 & 0.604±0.022 & 0.095±0.003 & 1680±60.73 & 92.2532/38 \\ 
        \textbf{} & ~ & 0-5\% & 3.4 & $<$ 0.05 & $10^2$ & 0.171±0.005 & 0.054±0.004 & 0.604±0.021 & 0.125±0.004 & 2300±130.35 & 65.6832/36 \\ 
        \textbf{} & ~ & 0-5\% & 7.7 & $<$ 0.1 & $10^3$ & 0.194±0.004 & 0.108±0.003 & 0.356±0.017 & 0.139±0.002 & 9260±333.36 & 15.5684/24 \\ 
        \textbf{} & ~ & 0-5\% & 11.5 & $<$ 0.1 & $10^4$ & 0.204±0.004 & 0.117±0.004 & 0.356±0.017 & 0.148±0.003 & 11360±408.96 & 16.5712/24 \\ 
        \textbf{Fig. 1 (b)} & $\pi^-$ & 0-5$\%$ & 14.5 & $<$ 0.1 & $10^5$ & 0.209±0.005 & 0.119±0.004 & 0.354±0.017 & 0.151±0.003 & 12861±462.996 & 14.9154/26 \\ 
        \textbf{Au-Au} & ~ & 0-5\% & 19.6 & $<$ 0.1 & $10^6$ & 0.209±0.005 & 0.129±0.004 & 0.354±0.016 & 0.157±0.002 & 14060±506.16 & 30.7145/24 \\ 
        \textbf{} & ~ & 0-5\% & 27 & $<$ 0.1 & $10^7$ & 0.214±0.004 & 0.127±0.005 & 0.354±0.017 & 0.157±0.003 & 15639±563.004 & 21.4635/24 \\ 
        \textbf{} & ~ & 0-5\% & 39 & $<$ 0.1 & $10^8$ & 0.223±0.005 & 0.122±0.004 & 0.354±0.017 & 0.157±0.003 & 16960±610.56 & 6.4209/24 \\ 
        \textbf{} & ~ & 0-10\% & 62.4 & $<$ 0.5 & $10^9$ & 0.24±0.005 & 0.112±0.004 & 0.354±0.018 & 0.157±0.004 & 23750±855 & 6.7172/11 \\ 
        \textbf{} & ~ & 0-5\% & 130 & $|\eta|$ $<$ 0.35 & $10^{10}$ & 0.24±0.006 & 0.112±0.005 & 0.354±0.015 & 0.157±0.003 & 24150±859 & 106.8680/12 \\ 
        \textbf{} & ~ & 0-5\% & 200 & $|\eta|$ $<$ 0.35 & $10^{11}$ & 0.253±0.006 & 0.11±0.004 & 0.334±0.016 & 0.157±0.003 & 29250±1053 & 50.7548/26 \\ 
        \hline
        \textbf{} & ~ & 0-5\% & 1.9 & $<$ 0.25 & $4\times10^2$ & 0.191±0.003 & 0.06±0.003 & 0.554±0.026 & 0.113±0.004 & 7.7E-4±0.00002 & 8.6280/7 \\ 
        \textbf{} & ~ & 0-5\% & 2.7 & $<$ 0.25 & $10^2$ & 0.270±0.004 & 0.060±0.003 & 0.454±0.022 & 0.155±0.005 & 0.063±0.007 & 1.9971/6 \\ 
        \textbf{} & ~ & 0-5\% & 3.4 & $<$ 0.25 & $10^2$ & 0.174±0.003 & 0.078±0.004 & 0.454±0.021 & 0.168±0.006 & 1.51±0.07 & 2.3082/3 \\ 
        \textbf{} & ~ & 0-5\% & 7.7 & $<$ 0.1 & 1 & 0.219±0.005 & 0.158±0.005 & 0.356±0.017 & 0.18±0.004 & 2060±74.16 & 7.6773/21 \\ 
        \textbf{} & ~ & 0-5\% & 11.5 & $<$ 0.1 & 10 & 0.221±0.005 & 0.164±0.006 & 0.356±0.018 & 0.184±0.004 & 2490±89.64 & 2.3743/23 \\ 
        \textbf{Fig. 1 (c)} & $K^+$ & 0-5\% & 14.5 & $<$ 0.1 & $10^2$ & 0.225±0.004 & 0.164±0.005 & 0.356±0.017 & 0.186±0.003 & 2699±97.164 & 4.8246/24 \\ 
        \textbf{Au-Au} & ~ & 0-5\% & 19.6 & $<$ 0.1 & $10^3$ & 0.234±0.005 & 0.167±0.005 & 0.356±0.017 & 0.191±0.004 & 2900±104.4 & 4.8785/24 \\ 
        \textbf{} & ~ & 0-5\% & 27 & $<$ 0.1 & $10^4$ & 0.234±0.006 & 0.167±0.005 & 0.356±0.016 & 0.191±0.004 & 3090±111.24 & 13.7093/24 \\ 
        \textbf{} & ~ & 0-5\% & 39 & $<$ 0.1 & $10^5$ & 0.245±0.005 & 0.161±0.004 & 0.356±0.017 & 0.191±0.005 & 3250±117 & 20.4385/24 \\ 
        \textbf{} & ~ & 0-5\% & 54.4 & $<$ 0.1 & $10^6$ & 0.265±0.006 & 0.15±0.005 & 0.356±0.017 & 0.191±0.004 & 3660±131.76 & 17.4267/21 \\ 
        \textbf{} & ~ & 0-5\% & 130 & $|\eta|$ $<$ 0.35 & $10^7$ & 0.265±0.006 & 0.15±0.005 & 0.356±0.017 & 0.191±0.004 & 5250±160.23 & 33.4002/11 \\ 
        \textbf{} & ~ & 0-5\% & 200 & $|\eta|$ $<$ 0.35 & $10^8$ & 0.294±0.006 & 0.134±0.004 & 0.356±0.017 & 0.191±0.004 & 5760±207.36 & 39.1350/14 \\  
        \hline
        \textbf{} & ~ & 0-5\% & 7.7 & $<$ 0.1 & 1 & 0.219±0.005 & 0.158±0.005 & 0.356±0.017 & 0.18±0.004 & 762±27.432 & 8.2908/21 \\ 
        \textbf{} & ~ & 0-5\% & 11.5 & $<$ 0.1 & 10 & 0.221±0.006 & 0.164±0.005 & 0.356±0.016 & 0.184±0.005 & 1260±45.36 & 7.6467/21 \\ 
        \textbf{} & ~ & 0-5\% & 14.5 & $<$ 0.1 & $10^2$ & 0.225±0.005 & 0.164±0.006 & 0.356±0.017 & 0.186±0.004 & 1580±56.88 & 26.9784/24 \\ 
        \textbf{Fig. 1 (d)} & $K^-$ & 0-5\% & 19.6 & $<$ 0.1 & $10^3$ & 0.234±0.005 & 0.167±0.005 & 0.356±0.017 & 0.191±0.004 & 1810±65.16 & 19.2545/24 \\ 
        \textbf{Au-Au} & ~ & 0-5\% & 27 & $<$ 0.1 & $10^4$ & 0.234±0.004 & 0.167±0.004 & 0.356±0.018 & 0.191±0.003 & 2190±78.84 & 16.2157/23 \\ 
        \textbf{} & ~ & 0-5\% & 39 & $<$ 0.1 & $10^5$ & 0.245±0.005 & 0.161±0.005 & 0.356±0.017 & 0.191±0.004 & 2560±92.16 & 16.3818/24 \\ 
        \textbf{} & ~ & 0-5\% & 130 & $|\eta|$ $<$ 0.35 & $10^6$ & 0.265±0.006 & 0.15±0.005 & 0.356±0.017 & 0.191±0.004 & 3650±98.23 & 9.9600/11 \\ 
        \textbf{} & ~ & 0-5\% & 200 & $|\eta|$ $<$ 0.35 & $10^7$ & 0.294±0.006 & 0.134±0.004 & 0.356±0.017 & 0.191±0.004 & 5360±192.96 & 54.8569/14 \\ 
        \hline
   \end{tabular}%
\end{center}
\end{sidewaystable}

\begin{sidewaystable}
\small Table 1 (to be continued...) \vspace{-.10cm}
\scriptsize
\begin{center}
    \begin{tabular}{cccccccccccccc}
    \hline
    \hline
     \textbf{Collision} & \textbf{Particle} & \textbf{C$\%$} & \textbf{Energy [GeV]} & \textbf{$|y|$} & \textbf{S.F} & \textbf{$T_1$ [GeV]} & \textbf{$T_2$ [GeV]} & \textbf{K} & \textbf{$T$ [GeV]} & \textbf{$N_0$} & \textbf{$\chi^2$/NDF} \\ 
        \hline
        \hline
        \textbf{} & ~ & $0-5\%$ & 1.9 & $<$ 0.05 & 1 & 0.251±0.003 & 0.049±0.002 & 0.454±0.022 & 0.141±0.003 & 1670±59.54 & 44.1985/37 \\ 
        \textbf{} & ~ & $0-5\%$ & 2.7 & $<$ 0.05 & 10 & 0.262±0.004 & 0.051±0.003 & 0.554±0.021 & 0.168±0.003 & 1650±50.76 & 78.2564/38 \\ 
        \textbf{} & ~ & $0-5\%$ & 3.4 & $<$ 0.05 & $10^2$ & 0.302±0.003 & 0.058±0.004 & 0.554±0.022 & 0.193±0.004 & 1550±45.32 & 66.0963/38 \\ 
        \textbf{} & ~ & $0-5\%$ & 7.7 & $<$ 0.1 & $10^3$ & 0.241±0.005 & 0.221±0.007 & 0.356±0.017 & 0.228±0.005 & 5560±200.16 & 13.2445/27 \\ 
        \textbf{} & ~ & $0-5\%$ & 11.5 & $<$ 0.1 & $10^4$ & 0.246±0.005 & 0.226±0.006 & 0.356±0.017 & 0.233±0.004 & 4560±164.16 & 24.0438/26 \\ 
        \textbf{Fig. 1 (e)} & $p$ & $0-5\%$ & 14.5 & $<$ 0.1 & $10^5$ & 0.251±0.006 & 0.231±0.007 & 0.356±0.018 & 0.238±0.005 & 3960±142.56 & 6.2376/23 \\ 
        \textbf{Au-Au} & ~ & $0-5\%$ & 19.6 & $<$ 0.1 & $10^6$ & 0.254±0.006 & 0.237±0.008 & 0.356±0.017 & 0.243±0.005 & 3550±127.8 & 9.5897/27 \\ 
        \textbf{} & ~ & $0-5\%$ & 27 & $<$ 0.1 & $10^7$ & 0.254±0.007 & 0.237±0.008 & 0.356±0.017 & 0.243±0.006 & 3150±113.4 & 6.3034/21 \\ 
        \textbf{} & ~ & $0-5\%$ & 39 & $<$ 0.1 & $10^8$ & 0.262±0.006 & 0.233±0.007 & 0.356±0.016 & 0.243±0.004 & 2801±100.836 & 24.3442/20 \\ 
        \textbf{} & ~ & $0-5\%$ & 54.4 & $<$ 0.1 & $10^9$ & 0.247±0.005 & 0.241±0.006 & 0.356±0.017 & 0.243±0.005 & 2901±104.436 & 26.6443/19 \\ 
        \textbf{} & ~ & $0-10\%$ & 62.4 & $<$ 0.5 & $10^{10}$ & 0.247±0.005 & 0.241±0.008 & 0.356±0.017 & 0.243±0.005 & 2761±99.396 & 73.2623/8 \\ 
        \textbf{} & ~ & $0-5\%$ & 130 & $|\eta|$ $<$ 0.35 & $10^{11}$ & 0.247±0.004 & 0.241±0.009 & 0.356±0.016 & 0.243±0.005 & 3250±100.25 & 30.1149/15 \\ 
        \textbf{} & ~ & $0-5\%$ & 200 & $|\eta|$ $<$ 0.35 & $10^{12}$ & 0.321±0.007 & 0.201±0.006 & 0.348±0.017 & 0.243±0.005 & 1690±60.84 & 87.0351/20 \\
        \hline
        \textbf{} & ~ & $0-5\%$ & 7.7 & $<$ 0.1 & 1 & 0.241±0.005 & 0.221±0.007 & 0.356±0.017 & 0.228±0.005 & 40±1.44 & 12.8654/13 \\ 
        \textbf{} & ~ & $0-5\%$ & 11.5 & $<$ 0.1 & 10 & 0.246±0.005 & 0.226±0.007 & 0.356±0.017 & 0.233±0.005 & 147±5.292 & 44.2003/21 \\ 
        \textbf{} & ~ & $0-5\%$ & 14.5 & $<$ 0.1 & $10^2$ & 0.251±0.006 & 0.231±0.007 & 0.356±0.016 & 0.238±0.006 & 250±9 & 16.8523/23 \\ 
        \textbf{} & ~& $0-5\%$ & 19.6 & $<$ 0.1 & $10^3$ & 0.254±0.006 & 0.237±0.008 & 0.356±0.017 & 0.243±0.004 &  430±15.48 & 15.8289/20 \\ 
        \textbf{Fig. 1 (f)} & $\Bar{p}$ & $0-5\%$ & 27 & $<$ 0.1 & $10^4$ & 0.254±0.006 & 0.237±0.008 & 0.356±0.017 & 0.243±0.005 & 650±23.4 & 24.5698/20 \\ 
        \textbf{Au-Au} & ~ & $0-5\%$ & 39 & $<$ 0.1 & $10^5$ & 0.262±0.006 & 0.233±0.007 & 0.356±0.018 & 0.243±0.005 & 871±31.356 & 43.4597/21 \\ 
        \textbf{} & ~ & $0-10\%$ & 62.4 & $<$ 0.5 & $10^6$ & 0.247±0.005 & 0.241±0.008 & 0.356±0.017 & 0.243±0.005 & 1161±41.796 & 64.8123/8 \\ 
        \textbf{} & ~ & $0-5\%$ & 130 & $|\eta|$ $<$ 0.35 & $10^7$ & 0.247±0.005 & 0.241±0.008 & 0.356±0.017 & 0.243±0.005 & 2250±70.74 & 49.8594/14 \\ 
        \textbf{} & ~ & $0-5\%$ & 200 & $|\eta|$ $<$ 0.35 & $10^8$ & 0.321±0.007 & 0.201±0.006 & 0.348±0.017 & 0.243±0.005 & 1260±45.36 & 120.9282/20 \\ 
        \hline
        \textbf{} & ~ & $0-5\%$ & 7.7 & $<$ 0.5 & 1 & 0.282±0.006 & 0.214±0.007 & 0.348±0.017 & 0.238±0.005 & 1700±61.2 & 45.5463/11 \\ 
        \textbf{} & ~ & $0-5\%$ & 11.5 & $<$ 0.5 & 10 & 0.292±0.005 & 0.214±0.007 & 0.428±0.021 & 0.248±0.005 & 1660±59.76 & 93.1509/12 \\ 
        \textbf{} & ~ & $0-5\%$ & 19.6 & $<$ 0.5 & $10^2$ & 0.277±0.007 & 0.225±0.006 & 0.848±0.042 & 0.269±0.006 & 1380±49.68 & 138.4923/13 \\ 
        \textbf{Fig. 1 (g)} & $\Lambda$ & $0-5\%$ & 27 & $<$ 0.5 & $10^3$ & 0.274±0.006 & 0.255±0.008 & 0.718±0.035 & 0.269±0.005 &  1210±43.56 & 68.6768/14 \\ 
        \textbf{Au-Au} & ~ & $0-5\%$ & 39 & $<$ 0.5 & $10^4$ & 0.284±0.006 & 0.229±0.007 & 0.721±0.034 & 0.269±0.005 & 1170±42.12 & 32.8943/14 \\ 
        \textbf{} & ~ & $0-5\%$ & 62.4 & $<$ 1 & $10^5$ & 0.284±0.006 & 0.229±0.007 & 0.721±0.035 & 0.269±0.004 & 2060±74.16 & 9.4215/10 \\ 
        \textbf{} & ~ & $0-5\%$ & 130 & $<$ 0.05 & $10^6$ & 0.284±0.006 & 0.229±0.007 & 0.721±0.035 & 0.269±0.004 & 8350±90.45 & 13.8818/8 \\ 
        \textbf{} & ~ & $0-5\%$ & 200 & $<$ 1 & $10^8$ & 0.326±0.007 & 0.129±0.004 & 0.711±0.035 & 0.269±0.005 & 1810±65.16 & 17.4503/15 \\ 
       \hline
        \textbf{} & ~ & $0-5\%$ & 7.7 & $<$ 0.5 & 1 & 0.282±0.006 & 0.214±0.007 & 0.348±0.017 & 0.238±0.005 & 21±0.756 & 21.9432/8 \\ 
        \textbf{} & ~ & $0-5\%$ & 11.5 & $<$ 0.5 & 10 & 0.292±0.006 & 0.214±0.007 & 0.428±0.022 & 0.248±0.005 & 75±2.7 & 80.8076/11 \\ 
        \textbf{} & ~ & $0-5\%$ & 19.6 & $<$ 0.5 & $10^2$ & 0.277±0.005 & 0.225±0.007 & 0.848±0.042 & 0.269±0.006 & 200±7.2 & 131.4524/12 \\ 
        \textbf{Fig. 1 (h)} & $\Bar{\Lambda}$ & $0-5\%$ & 27 & $<$ 0.5 & $10^3$ & 0.274±0.007 & 0.255±0.008 & 0.718±0.035 & 0.269±0.005  & 310±11.16 & 98.3465/13 \\ 
        \textbf{Au-Au} & ~ & $0-5\%$ & 39 & $<$ 0.5 & $10^4$ & 0.284±0.006 & 0.229±0.007 & 0.721±0.035 & 0.269±0.005 & 381±13.716 & 45.5753/14 \\ 
        \textbf{} & ~ & $0-5\%$ & 62.4 & $<$ 1 & $10^5$ & 0.284±0.006 & 0.229±0.007 & 0.721±0.034 & 0.269±0.004 & 1020±36.72 & 69.1245/10 \\ 
        \textbf{} & ~ & $0-5\%$ & 130 & $<$ 0.05 & $10^6$ & 0.284±0.006 & 0.229±0.007 & 0.721±0.035 & 0.269±0.004 & 840±36.68 & 37.7712/8 \\ 
        \textbf{} & ~ & $0-5\%$ & 200 & $<$ 1 & $10^7$ & 0.326±0.007 & 0.129±0.004 & 0.711±0.035 & 0.269±0.005 & 1310±47.16 & 16.9300/15 \\ 
         \hline
         \textbf{} & ~ & $0-5\%$ & 7.7 & $<$ 0.5 & 1 & 0.281±0.006 & 0.225±0.007 & 0.348±0.016 & 0.245±0.005 & 124±4.464 & 4.6373/5 \\ 
        \textbf{} & ~ & $0-5\%$ & 11.5 & $<$ 0.5 & 10 & 0.289±0.006 & 0.225±0.007 & 0.398±0.02 & 0.251±0.005 & 163±5.868 & 20.6342/6 \\ 
        \textbf{} & ~ & $0-5\%$ & 19.6 & $<$ 0.5 & $10^2$ & 0.318±0.007 & 0.255±0.008 & 0.398±0.02 & 0.28±0.006 & 168±6.048 & 2.7458/6 \\ 
        \textbf{Fig. 1 (i)} & $\Xi^-$ & $0-5\%$ & 27 & $<$ 0.5 & $10^3$ & 0.318±0.008 & 0.255±0.008 & 0.398±0.02 & 0.28±0.005 & 184±6.624 & 16.4476/7 \\ 
        \textbf{Au-Au} & ~ & $0-5\%$ & 39 & $<$ 0.5 & $10^4$ & 0.318±0.007 & 0.255±0.008 & 0.398±0.01 & 0.28±0.006 & 190±6.84 & 8.9434/7 \\ 
        \textbf{} & ~ & $0-5\%$ & 62.4 & $<$ 1 & $10^5$ & 0.3±0.007 & 0.266±0.009 & 0.418±0.02 & 0.28±0.006 & 199±7.164 & 52.2694/9 \\ 
        \textbf{} & ~ & $0-10\%$ & 130 & $<$ 0.05 & $10^6$ & 0.3±0.006 & 0.266±0.007 & 0.418±0.02 & 0.28±0.005 & 189±6.123 & 6.2038/8 \\ 
        \textbf{} & ~ & $0-5\%$ & 200 & $<$ 0.75 & $10^7$ & 0.339±0.007 & 0.141±0.005 & 0.701±0.034 & 0.28±0.005 & 257±9.252 & 11.0395/13 \\ 
       \hline
   \end{tabular}%
\end{center}
\end{sidewaystable}

\begin{sidewaystable}
\small Table 1 (to be continued...) \vspace{-.10cm}
\scriptsize
\begin{center}
    \begin{tabular}{cccccccccccccc}
    \hline
    \hline
     \textbf{Collision} & \textbf{Particle} & \textbf{C$\%$} & \textbf{Energy [GeV]} & \textbf{$|y|$} & \textbf{S.F} & \textbf{$T_1$ [GeV]} & \textbf{$T_2$ [GeV]} & \textbf{K} & \textbf{$T$ [GeV]} & \textbf{$N_0$} & \textbf{$\chi^2$/NDF} \\ 
        \hline
        \hline
         \textbf{} & ~ & $0-5\%$ & 7.7 & $<$ 0.5 & 1 & 0.281±0.006 & 0.225±0.007 & 0.348±0.017 & 0.245±0.005 & 7.2±0.259 & 1.7453/4 \\ 
        \textbf{} & ~ & $0-5\%$ & 11.5 & $<$ 0.5 & 10 & 0.289±0.006 & 0.225±0.007 & 0.398±0.02 & 0.251±0.005 & 20±0.72 & 15.1094/6 \\ 
        \textbf{} & ~ & $0-5\%$ & 19.6 & $<$ 0.5 & $10^2$ & 0.318±0.007 & 0.255±0.008 & 0.398±0.03 & 0.28±0.006 & 46±1.656 & 27.4925/6 \\ 
        \textbf{Fig. 1 (j)} & $\Bar{\Xi^+}$ & $0-5\%$ & 27 & $<$ 0.5 & $10^3$ & 0.318±0.008 & 0.255±0.008 & 0.398±0.01 & 0.28±0.005 & 69±2.484 & 35.0656/7 \\ 
        \textbf{Au-Au} & ~ & $0-5\%$ & 39 & $<$ 0.5 & $10^4$ & 0.318±0.007 & 0.255±0.008 & 0.398±0.02 & 0.28±0.006 & 92±3.312 & 11.3337/7 \\ 
        \textbf{} & ~ & $0-5\%$ & 62.4 & $<$ 1 & $10^5$ & 0.3±0.007 & 0.266±0.009 & 0.418±0.02 & 0.28±0.006 & 130±4.68 & 32.0486/9 \\ 
        \textbf{} & ~ & $0-10\%$ & 130 & $<$ 0.05 & $10^6$ & 0.3±0.006 & 0.266±0.008 & 0.418±0.02 & 0.28±0.005 & 143±6.76 & 11.1093/8 \\ 
        \textbf{} & ~ & $0-5\%$ & 200 & $<$ 0.75 & $10^7$ & 0.339±0.007 & 0.141±0.005 & 0.701±0.033 & 0.28±0.005 & 209±7.524 & 6.4563/13 \\
        \hline
   \end{tabular}%
\end{center}
\end{sidewaystable}

\appendix
\begin{table}
\small \textbf{Table 2.} The values of $\langle p_T \rangle$ calculated from the probability density function i.e., calculated from Eq. (11) and $T_i$ calculated from the string Percolation model given in Eq. (13). 
\scriptsize
\begin{center}
    \begin{tabular}{ccccccc}
    \hline
    \hline
        \textbf{Collision} & \textbf{Particle} & \textbf{C$\%$} & \textbf{Energy [GeV]} & \textbf{$|y|$} & \textbf{$\langle p_T \rangle$ [GeV/c]} & \textbf{$T_i$ [GeV]} \\ 
        \hline
        \hline
        \textbf~ & $~$ &  0-5\% & 1.9 & $<$ 0.05 & 0.228±0.005 & 0.201±0.007 \\ 
        \textbf{} & ~ &  0-5\% & 2.7 & $<$ 0.05 & 0.276±0.008 & 0.243±0.011 \\ 
        \textbf{} & ~ & 0-5\% & 3.4 & $<$ 0.05 & 0.356±0.010 & 0.323±0.013 \\ 
        \textbf{} & ~ & 0-5\% & 7.7 & $<$ 0.1 & 0.391±0.013 & 0.338±0.012 \\ 
        \textbf{} & ~ & 0-5\% & 11.5 & $<$ 0.1 & 0.416±0.013 & 0.359±0.012 \\ 
        \textbf{Au-Au} & $\pi^+$ & 0-5\% & 14.5 & $<$ 0.1 & 0.423±0.014 & 0.366±0.011 \\ 
        \textbf{} & ~ & 0-5\% & 19.6 & $<$ 0.1 & 0.44±0.014 & 0.377±0.013 \\ 
        \textbf{} & ~ & 0-5\% & 27 & $<$ 0.1 & 0.44±0.015 & 0.379±0.013 \\ 
        \textbf{} & ~ & 0-5\% & 39 & $<$ 0.1 & 0.44±0.014 & 0.383±0.013 \\ 
        \textbf{} & ~ & 0-5\% & 54.4 & $<$ 0.1 & 0.44±0.014 & 0.383±0.014 \\ 
        \textbf{} & ~ & 0-10\% & 62.4 & $<$ 0.5 & 0.426±0.014 & 0.379±0.013 \\ 
        \textbf{} & ~ & 0-5\% & 130 & $|\eta|$ $<$ 0.35 & 0.433±0.016 & 0.386±0.015 \\ 
        \textbf{} & ~ & 0-5\% & 200 & $|\eta|$ $<$ 0.35 & 0.433±0.014 & 0.391±0.013 \\ 
        \hline
        \textbf~ & ~ & 0-5\% & 1.9 & $<$ 0.05 & 0.228±0.005 & 0.201±0.007 \\ 
        \textbf{} & ~ & 0-5\% & 2.7 & $<$ 0.05 & 0.276±0.008 & 0.243±0.011 \\ 
        \textbf{} & ~ & 0-5\% & 3.4 & $<$ 0.05 & 0.356±0.010 & 0.323±0.013 \\ 
        \textbf{} & ~ & 0-5\% & 7.7 & $<$ 0.1 & 0.391±0.013 & 0.338±0.012 \\ 
        \textbf{} & ~ & 0-5\% & 11.5 & $<$ 0.1 & 0.416±0.013 & 0.359±0.011 \\ 
        \textbf{Au-Au} & $\pi^-$ & 0-5$\%$ & 14.5 & $<$ 0.1 & 0.423±0.014 & 0.366±0.012 \\ 
        \textbf{} & ~ & 0-5\% & 19.6 & $<$ 0.1 & 0.44±0.014 & 0.377±0.013 \\ 
        \textbf{} & ~ & 0-5\% & 27 & $<$ 0.1 & 0.44±0.013 & 0.379±0.013 \\ 
        \textbf{} & ~ & 0-5\% & 39 & $<$ 0.1 & 0.44±0.014 & 0.383±0.014 \\ 
        \textbf{} & ~ & 0-10\% & 62.4 & $<$ 0.5 & 0.426±0.015 & 0.379±0.013\\ 
        \textbf{} & ~ & 0-5\% & 130 & $|\eta|$ $<$ 0.35 & 0.433±0.016 & 0.386±0.015 \\ 
        \textbf{} & ~ & 0-5\% & 200 & $|\eta|$ $<$ 0.35 & 0.433±0.014 & 0.391±0.013 \\ 
        \hline
        \textbf{} & ~ & 0-5\% & 1.9 & $<$ 0.25 & 0.441±0.006 & 0.388±0.009 \\ 
        \textbf{} & ~ & 0-5\% & 2.7 & $<$ 0.25 & 0.500±0.011 & 0.470±0.013 \\ 
        \textbf{} & ~ & 0-5\% & 3.4 & $<$ 0.25 & 0.532±0.013 & 0.486±0.014 \\ 
        \textbf{} & ~ & 0-5\% & 7.7 & $<$ 0.1  & 0.568±0.018 & 0.471±0.016 \\ 
        \textbf{} & ~ & 0-5\% & 11.5 & $<$ 0.1 & 0.58±0.019 & 0.481±0.016 \\ 
        \textbf{Au-Au} & $K^+$ & 0-5\% & 14.5 & $<$ 0.1 & 0.584±0.018 & 0.484±0.016 \\ 
        \textbf{} & ~ & 0-5\% & 19.6 & $<$ 0.1 & 0.597±0.019 & 0.496±0.017 \\ 
        \textbf{} & ~ & 0-5\% & 27 & $<$ 0.1 & 0.597±0.019 & 0.496±0.017 \\ 
        \textbf{} & ~ & 0-5\% & 39 & $<$ 0.1 & 0.597±0.020 & 0.499±0.018 \\ 
        \textbf{} & ~ & 0-5\% & 54.4 & $<$ 0.1 & 0.598±0.019 & 0.507±0.017 \\ 
        \textbf{} & ~ & 0-5\% & 130 & $|\eta|$ $<$ 0.35 & 0.589±0.018 & 0.499±0.016 \\ 
        \textbf{} & ~ & 0-5\% & 200 & $|\eta|$ $<$ 0.35 & 0.512±0.016 & 0.589±0.02 \\  
        \hline
        \textbf{} & ~ & 0-5\% & 7.7 & $<$ 0.1 & 0.568±0.018 & 0.471±0.016 \\ 
        \textbf{} & ~ & 0-5\% & 11.5 & $<$ 0.1 & 0.58±0.019 & 0.481±0.015 \\ 
        \textbf{} & ~ & 0-5\% & 14.5 & $<$ 0.1 & 0.584±0.019 & 0.484±0.016 \\ 
        \textbf{Au-Au} & $K^-$ & 0-5\% & 19.6 & $<$ 0.1 & 0.597±0.020 & 0.496±0.017 \\ 
        \textbf{} & ~ & 0-5\% & 27 & $<$ 0.1 & 0.597±0.019 & 0.496±0.018 \\ 
        \textbf{} & ~ & 0-5\% & 39 & $<$ 0.1 & 0.597±0.019 & 0.499±0.017\\ 
        \textbf{} & ~ & 0-5\% & 130 & $|\eta|$ $<$ 0.35 & 0.589±0.018 & 0.499±0.016 \\ 
        \textbf{} & ~ & 0-5\% & 200 & $|\eta|$ $<$ 0.35 & 0.512±0.016 & 0.589±0.02 \\ 
        \hline
        \textbf{} & ~ & $0-5\%$ & 1.9 & $<$ 0.05 & 0.544±0.012 & 0.499±0.013 \\ 
        \textbf{} & ~ & $0-5\%$ & 2.7 & $<$ 0.05 & 0.620±0.013 & 0.559±0.014 \\ 
        \textbf{} & ~ & $0-5\%$ & 3.4 & $<$ 0.05 & 0.691±0.014 & 0.628±0.016 \\ 
        \textbf{} & ~ & $0-5\%$ & 7.7 & $<$ 0.1 & 0.788±0.025 & 0.643±0.022 \\ 
        \textbf{} & ~ & $0-5\%$ & 11.5 & $<$ 0.1 & 0.801±0.026 & 0.654±0.022 \\ 
        \textbf{Au-Au} & $p$ & $0-5\%$ & 14.5 & $<$ 0.1 & 0.815±0.026 & 0.665±0.023 \\ 
        \textbf{} & ~ & $0-5\%$ & 19.6 & $<$ 0.1 & 0.829±0.027 & 0.677±0.024 \\ 
        \textbf{} & ~ & $0-5\%$ & 27 & $<$ 0.1 & 0.829±0.027 & 0.677±0.022 \\ 
        \textbf{} & ~ & $0-5\%$ & 39 & $<$ 0.1 & 0.829±0.028 & 0.677±0.023 \\ 
        \textbf{} & ~ & $0-5\%$ & 54.4 & $<$ 0.1 & 0.829±0.027 & 0.676±0.023 \\ 
        \textbf{} & ~ & $0-10\%$ & 62.4 & $<$ 0.5 & 0.807±0.026 & 0.659±0.022 \\ 
        \textbf{} & ~ & $0-5\%$ & 130 & $|\eta|$ $<$ 0.35 & 0.818±0.028 & 0.668±0.024 \\ 
        \textbf{} & ~ & $0-5\%$ & 200 & $|\eta|$ $<$ 0.35 & 0.819±0.026 & 0.68±0.023 \\
        \hline
        \textbf{} & ~ & $0-5\%$ & 7.7 & $<$ 0.1 & 0.788±0.025 & 0.643±0.022 \\ 
        \textbf{} & ~ & $0-5\%$ & 11.5 & $<$ 0.1 & 0.801±0.026 & 0.654±0.022 \\ 
        \textbf{} & ~ & $0-5\%$ & 14.5 & $<$ 0.1 & 0.815±0.026 & 0.665±0.023 \\ 
        \textbf{} & ~& $0-5\%$ & 19.6 & $<$ 0.1 & 0.829±0.027 & 0.677±0.023 \\ 
        \textbf{Au-Au} & $\Bar{p}$ & $0-5\%$ & 27 & $<$ 0.1 & 0.829±0.027 & 0.677±0.022 \\ 
        \textbf{} & ~ & $0-5\%$ & 39 & $<$ 0.1 & 0.829±0.027 & 0.677±0.024 \\ 
        \textbf{} & ~ & $0-10\%$ & 62.4 & $<$ 0.5 & 0.807±0.026 & 0.659±0.023 \\ 
        \textbf{} & ~ & $0-5\%$ & 130 & $|\eta|$ $<$ 0.35 & 0.818±0.028 & 0.668±0.024 \\ 
        \textbf{} & ~ & $0-5\%$ & 200 & $|\eta|$ $<$ 0.35 & 0.819±0.026 & 0.68±0.023 \\ 
        \hline
   \end{tabular}%
\end{center}
\end{table}

\begin{table}
\small Table 2 (to be continued...) \vspace{-.10cm}
\scriptsize
\begin{center}
    \begin{tabular}{cccccccccccccc}
    \hline
    \hline
     \textbf{Collision} & \textbf{Particle} & \textbf{C$\%$} & \textbf{Energy [GeV]} & \textbf{$|y|$} & \textbf{$\langle p_T \rangle$ [GeV/c]} & \textbf{$T_i$ [GeV]} \\ 
        \hline
        \hline
        \textbf{} & ~ & $0-5\%$ & 7.7 & $<$ 0.5  & 0.828±0.026 & 0.678±0.023 \\ 
        \textbf{} & ~ & $0-5\%$ & 11.5 & $<$ 0.5 & 0.853±0.027 & 0.7±0.024 \\ 
        \textbf{} & ~ & $0-5\%$ & 19.6 & $<$ 0.5 & 0.909±0.029 & 0.742±0.025 \\ 
        \textbf{Au-Au} & $\Lambda$ & $0-5\%$ & 27 & $<$ 0.5 & 0.909±0.030 & 0.741±0.024 \\ 
        \textbf{} & ~ & $0-5\%$ & 39 & $<$ 0.5 & 0.909±0.029 & 0.743±0.025 \\ 
        \textbf{} & ~ & $0-5\%$ & 62.4 & $<$ 1 & 0.879±0.028 & 0.719±0.024 \\ 
        \textbf{} & ~ & $0-5\%$ & 130 & $<$ 0.05 & 0.932±0.030 & 0.762±0.026 \\ 
        \textbf{} & ~ & $0-5\%$ & 200 & $<$ 1 & 0.878±0.028 & 0.739±0.025 \\ 
       \hline
        \textbf{} & ~ & $0-5\%$ & 7.7 & $<$ 0.5 & 0.828±0.026 & 0.678±0.023 \\ 
        \textbf{} & ~ & $0-5\%$ & 11.5 & $<$ 0.5 & 0.853±0.027 & 0.7±0.024 \\ 
        \textbf{} & ~ & $0-5\%$ & 19.6 & $<$ 0.5 & 0.909±0.029 & 0.742±0.025 \\ 
        \textbf{Au-Au} & $\Bar{\Lambda}$ & $0-5\%$ & 27 & $<$ 0.5 & 0.909±0.028 & 0.741±0.025 \\ 
        \textbf{} & ~ & $0-5\%$ & 39 & $<$ 0.5 & 0.909±0.029 & 0.743±0.025 \\ 
        \textbf{} & ~ & $0-5\%$ & 62.4 & $<$ 1 & 0.879±0.028 & 0.719±0.024 \\ 
        \textbf{} & ~ & $0-5\%$ & 130 & $<$ 0.05 & 0.932±0.030 & 0.762±0.026 \\ 
        \textbf{} & ~ & $0-5\%$ & 200 & $<$ 1 & 0.878±0.027 & 0.739±0.025 \\ 
         \hline
         \textbf{} & ~ & $0-5\%$ & 7.7 & $<$ 0.5 & 0.886±0.028 & 0.722±0.025 \\ 
        \textbf{} & ~ & $0-5\%$ & 11.5 & $<$ 0.5 & 0.901±0.029 & 0.736±0.025 \\ 
        \textbf{} & ~ & $0-5\%$ & 19.6 & $<$ 0.5 & 0.977±0.031 & 0.798±0.027 \\ 
        \textbf{Au-Au} & $\Xi^-$ & $0-5\%$ & 27 & $<$ 0.5 & 0.977±0.030 & 0.798±0.027 \\ 
        \textbf{} & ~ & $0-5\%$ & 39 & $<$ 0.5 & 0.977±0.031 & 0.798±0.027 \\ 
        \textbf{} & ~ & $0-5\%$ & 62.4 & $<$ 1 & 0.949±0.03 & 0.774±0.026 \\ 
        \textbf{} & ~ & $0-10\%$ & 130 & $<$ 0.05 & 1.002±0.04 & 0.816±0.030 \\ 
        \textbf{} & ~ & $0-5\%$ & 200 & $<$ 0.75 & 0.957±0.031 & 0.803±0.027 \\ 
       \hline
        \textbf{} & ~ & $0-5\%$ & 7.7 & $<$ 0.5 & 0.886±0.028 & 0.722±0.025 \\ 
        \textbf{} & ~ & $0-5\%$ & 11.5 & $<$ 0.5 & 0.901±0.029 & 0.736±0.025 \\ 
        \textbf{} & ~ & $0-5\%$ & 19.6 & $<$ 0.5 & 0.977±0.031 & 0.798±0.027 \\ 
        \textbf{Au-Au} & $\Bar{\Xi^+}$ & $0-5\%$ & 27 & $<$ 0.5 & 0.977±0.032 & 0.798±0.027 \\ 
        \textbf{} & ~ & $0-5\%$ & 39 & $<$ 0.5 & 0.977±0.031 & 0.798±0.027 \\ 
        \textbf{} & ~ & $0-5\%$ & 62.4 & $<$ 1 & 0.949±0.03 & 0.774±0.026 \\ 
        \textbf{} & ~ & $0-10\%$ & 130 & $<$ 0.05 & 1.002±0.04 & 0.816±0.030 \\ 
        \textbf{} & ~ & $0-5\%$ & 200 & $<$ 0.75 & 0.957±0.031 & 0.803±0.027 \\
        \hline
        \hline
   \end{tabular}%
\end{center}
\end{table}

\end{document}